\documentclass[twocolumn,showpacs,preprintnumbers,amsmath,amssymb,aps,prb,superscriptaddress]{revtex4-2}

\pdfoutput=1
\usepackage[pdftex]{graphicx}
\usepackage[english]{babel}
\usepackage{soul}
\usepackage{cleveref}
\usepackage{bbold}
\usepackage{mathtools}
\usepackage{multirow}
\usepackage{colortbl}
\definecolor{kugray5}{RGB}{224,224,224}
\usepackage[latin1]{inputenc}
\usepackage{diagbox}
\usepackage{mciteplus}
\usepackage{mathrsfs}

\usepackage{dcolumn}
\usepackage{bm}

\usepackage{color}
\usepackage{amstext}
\usepackage{braket}
\usepackage{mathdots}
\usepackage{tikz}
\usepackage{physics}
\usepackage{enumitem}
\usepackage{siunitx}

\usetikzlibrary{matrix,decorations.pathreplacing}

\begin{document}


\title{Impurity flat band states in the diamond chain}

\author{A. M. Marques}
\email{anselmomagalhaes@ua.pt}
\affiliation{Department of physics $\&$ i3N, University of Aveiro, 3810-193 Aveiro, Portugal}

\author{D. Viedma}
\affiliation{Departament de F\'isica, Universitat Aut\`onoma de Barcelona, E-08193 Bellaterra, Spain}

\author{V. Ahufinger}
\affiliation{Departament de F\'isica, Universitat Aut\`onoma de Barcelona, E-08193 Bellaterra, Spain}

\author{R. G. Dias}
\affiliation{Department of physics $\&$ i3N, University of Aveiro, 3810-193 Aveiro, Portugal}


\begin{abstract}
Flat band (FB) systems, featuring dispersionless energy bands, have garnered significant interest due to their compact localized states (CLSs). 
However, a detailed account on how local impurities affect the physical properties of overlapping CLSs is still missing. Here we study a diamond chain with a finite magnetic flux per plaquette that exhibits a gapped midspectrum FB with non-orthogonal CLSs, and develop a framework for projecting operators onto such non-orthogonal bases. 
This framework is applied to the case of an open diamond chain with small local impurities in the midchain plaquette, and analytical expressions are derived for FB states influenced by these impurities. For equal impurities in top and bottom sites under diagonal disorder, we show how the impurity states experience an averaged disorder dependent on their spatial extension, leading to enhanced robustness against disorder. 
For a single impurity, an exotic topological phase with a half-integer winding number is discovered, which is linked to a single in-gap edge state under open boundary conditions. 
Numerical simulations validate the analytical predictions.
\end{abstract}


\maketitle

\section*{Introduction}
\label{sec:intro}
About a decade ago, a renewed interest in flat band (FB) systems \cite{Flach2014} gave way to a new cycle of studies devoted to the topic (a more detailed chronology can be found in \cite{Danieli2024}), and it has no signs of slowing down.
In this recent wave of studies, the focus has gradually shifted towards the compact localized states (CLSs) that span most types of FBs.
For one-dimensional (1D) models, it has been shown that these CLSs form an orthogonal set when the FB projector is strictly local, and a non-orthogonal one otherwise \cite{Sathe2021,Sathe2024}.
Based on the form of the CLSs, FB generators were constructed for 1D \cite{Maimaiti2017,Maimaiti2019} and higher-dimensional \cite{Dias2015,Maimaiti2021,Hwang2021,Graf2021,Kim2023} models.

An interesting class of FB systems is the (self-explanatory) class of all-bands-flat (ABF) models \cite{Danieli2021,Li2020}, where transport is completely suppressed, since any local excitation in an ABF model can be written as a linear combination of CLSs belonging to the different FBs, leading to the phenomenon of particle confinement known as Aharonov-Bohm (AB) caging \cite{Mukherjee2020}.
The diamond chain with a $\pi$-flux per plaquette \cite{Mukherjee2018,Pelegri2019,Pelegri2019b,Liberto2019,Kremer2020,Aravena2022}, along with its $2^n$-root versions \cite{Marques2021,Marques2023}, is a prime example of an ABF model exhibiting AB caging that has been thoroughly investigated.
In this and other related ABF models, the effects of weak perturbations have been studied by projecting the perturbation operator onto one of the FBs.
For example, this method was employed to study weak disorder in the ABF Creutz ladder \cite{Cadez2021,Kim2023a}, and weak interactions in both this model \cite{Kuno2020b} and the diamond chain \cite{Pelegri2020}, shown to be able to induce dynamical effects \cite{Tovmasyan2013} in two-body states where it would otherwise be absent, due to the infinite effective mass of the states of a FB.
Conversely, strong interactions can drive the appearance of effective ABF many-body subspaces, in models with a dispersive single-particle energy spectrum \cite{Nicolau2023}.

Here, we address the case of a diamond chain with a finite flux per plaquette, and introduce a perturbation in the form of weak local potentials in the midchain plaquette, as schematically illustrated in Fig.~\ref{fig:chain}(a).
By projecting this impurity perturbation onto the gapped zero-energy FB, only two partially overlapping CLSs have finite weight at the impurity sites.
However, due to the non-orthogonal nature of adjacent CLSs, we show that the analytical treatment of the system requires one to reexpress the problem in terms of its dual basis states.
This greatly simplifies the problem, by reducing it to the solutions of a decoupled dimer in a non-orthogonal dual subspace.
We then derive the analytical solutions for the impurity states that are lifted from the FB as a function of the magnetic flux, focusing on the cases of two equal impurities, two opposite impurities, and a single-impurity, where near-perfect agreement with numerical results is observed.
Finally, for the single-impurity case, we further demonstrate how it corresponds to a non-trivial topological phase characterized by half-integer topological invariants.
Upon mapping this system onto an equivalent 2D lattice decorated with a central line of defects, we provide indirect evidence of a refined bulk-boundary correspondence by relating the half-integer bulk invariants with the presence of a single in-gap edge state in the spectrum of the mapped lattice.


\section*{Results}

\subsection*{Basics of non-orthogonal bases}
\label{sec:basics}
In this section, we will review some basic results of non-orthogonal bases, particularly in what concerns the definition of projection operators and the procedure for diagonalizing a Hamiltonian written in such a basis.
We will heavily rely in the theory outlined in \cite{Soriano2014}, as well in the summarized results provided in the appendix of \cite{Santos2020}.
Let us start by considering a non-orthogonal basis $\{\ket{i}\}$, not necessarily normalized.
The elements of the overlap matrix $\mathbf{S}$ of this basis are written as $S_{ij}=\braket{i}{j}$, with $\mathbf{S}$ further assumed to be a real symmetric matrix $S_{ij}=S_{ji}$.
Evidently, $\mathbf{S}$ reduces to the identity $\mathbf{I}$ for an orthonormal basis.
One can define the dual basis $\{\ket{i^*}\}$, whose states are defined by the property $\braket{i}{j^*}=\delta_{ij}$, with $\delta_{ij}$ the Kronecker delta.
The transformation of states between the direct and dual bases is given by
\begin{equation}
	\ket{i}=\sum\limits_j S_{ij}\ket{j^*},\ \ \ \ \ket{i^*}=\sum\limits_j S_{ij}^{-1}\ket{j},
	\label{eq:directdualtransf}
\end{equation}
where the usual slight abuse of language of writing $(\mathbf{S}^{-1})_{ij}$ as $S_{ij}^{-1}$ is used here and throughout the text for simplicity.
The identity operator can therefore be expressed in several different forms, as
\begin{equation}
	\hat{I}=\sum\limits_i\ket{i}\bra{i^*}=\sum\limits_{ij}\ket{i}S_{ij}^{-1}\bra{j},
\end{equation}
which is widely used, e.g., in systems of coupled coherent states \cite{Shalashilin2004,Symonds2015} that form an overcomplete non-orthogonal basis, or alternatively as
\begin{equation}
	\hat{I}=\sum\limits_i\ket{i^*}\bra{i}=\sum\limits_{ij}\ket{i^*}S_{ij}\bra{j^*}.
\end{equation}
From $\hat{I}^2=\hat{I}$, one can derive the following useful identity,
\begin{equation}
	\sum\limits_kS_{ik}^{-1}S_{kj}=\delta_{ij}.
	\label{eq:deltakron}
\end{equation}

\subsubsection*{Projection of operators onto a non-orthogonal subspace}
It is convenient, for later use, to consider the case of a basis of size $N+M$ composed of two orthogonal subspaces, $\{\ket{i}\}=\{\ket{\text{Loc},n}\}\cup\{\ket{\text{Ext},m}\}$, with $n=1,2,\dots,N$ and $m=1,2,\dots,M$, and where, anticipating the notation that will be used later on, $\{\ket{\text{Loc},n}\}$ stands for the non-orthogonal subspace of localized states and $\{\ket{\text{Ext},m}\}$ for the orthonormal subspace of extended states.
The total overlap matrix reads as $\mathbf{S}_{\text{tot}}=\mathbf{S}\oplus\mathbf{I}_M$, where $\mathbf{I}_M$ is the identity matrix of size $M$.
The projection operator onto the non-orthogonal subspace is written as
\begin{eqnarray}
	\hat{P}_{\text{Loc}}&=&\sum\limits_{j=1}^N\ket{\text{Loc},j}\bra{\text{Loc},j^*} \nonumber
	\\
	&=&\sum\limits_{j=1}^N\ket{\text{Loc},j}\sum\limits_{i=1}^{N+M}S_{ij}^{-1}\bra{i}
	\nonumber
	\\
	&=&\sum\limits_{i,j=1}^N\ket{\text{Loc},j}S_{ij}^{-1}\bra{\text{Loc},i},
\end{eqnarray}
where $\ket{\text{Loc},j^*}$ belongs to the dual localized subspace and the orthogonality between subspaces was used in the last line.
Correspondingly,
\begin{eqnarray}
	\hat{P}_{\text{Loc}}^\dagger&=&\sum\limits_{j=1}^N\ket{\text{Loc},j^*}\bra{\text{Loc},j} \nonumber\\
	&=&\sum\limits_{i,j=1}^N\ket{\text{Loc},i}S_{ij}^{-1}\bra{\text{Loc},j}=\hat{P}_{\text{Loc}}.
\end{eqnarray}
Given an arbitrary operator $\hat{A}$, its projection onto the non-orthogonal subspace can be developed as
\begin{eqnarray}
	\hat{A}_{\text{Loc}}&=&\hat{P}_{\text{Loc}}^\dagger\hat{A}\hat{P}_{\text{Loc}} \nonumber
	\\
	&=&\sum\limits_{i,j=1}^N\ket{\text{Loc},i^*}A_{ij}\bra{\text{Loc},j^*},
	\label{eq:projoperator}
\end{eqnarray}
where $A_{ij}=\bra{\text{Loc},i}\hat{A}\ket{\text{Loc},j}$.

\subsubsection*{Symmetric orthonormalization}
\label{subsec:lowdin}
Given a non-orthogonal basis, one can always find an orthonormalized basis through the Gram-Schmidt method \cite{Schmidt1907}, which relies on a recursive state-by-state orthogonalization procedure.
However, this method can be very computationally demanding and prone to numerical instabilities.
A more efficient method consists of using the L\"owdin transformation \cite{Loewdin1956,Leschber1989} to perform a symmetrical orthogonalization of a non-orthogonal basis.
With this transformation, and from a non-orthogonal basis spanning $\{\ket{i}\}$, with $i=1,2,\dots,N$, one can find the orthogonal basis states with the help of the overlap matrix as
\begin{equation}
	\ket{\tilde{i}}=\sum\limits_j S_{ij}^{-\frac{1}{2}}\ket{j},
	\label{eq:lowdin}
\end{equation} 
where, again, we use the simplified notation $S_{ij}^{-\frac{1}{2}}\equiv (\mathbf{S}^{-\frac{1}{2}})_{ij}$.
The secular equation, in this basis, is computed through the effective Hamiltonian $\tilde{\mathbf{H}}$,
\begin{eqnarray}
	\tilde{\mathbf{H}}\tilde{\pmb{\psi}_j}&=&E_j\tilde{\pmb{\psi}_j}, \ \ j=1,2,\dots,N,
	\label{eq:secularqe}
	\\
	\tilde{\mathbf{H}}&=&\mathbf{S}^{-\frac{1}{2}}\mathbf{H}\mathbf{S}^{-\frac{1}{2}},
	\label{eq:effhamilt}
\end{eqnarray}
where $\mathbf{H}$ is the Hamiltonian matrix written in the starting non-orthogonal $\{\ket{i}\}$ basis, also called the core Hamiltonian \cite{Suzuki1973}, and $\tilde{\pmb{\psi}_j}$ is the $j^{\text{th}}$ eigenvector with components spanning the orthogonal $\{\ket{\tilde{i}}\}$ basis.

\subsection*{Diamond chain with flux}
\label{sec:diamondchain}

\begin{figure}[ht]
	\begin{centering}
		\includegraphics[width=0.95 \columnwidth]{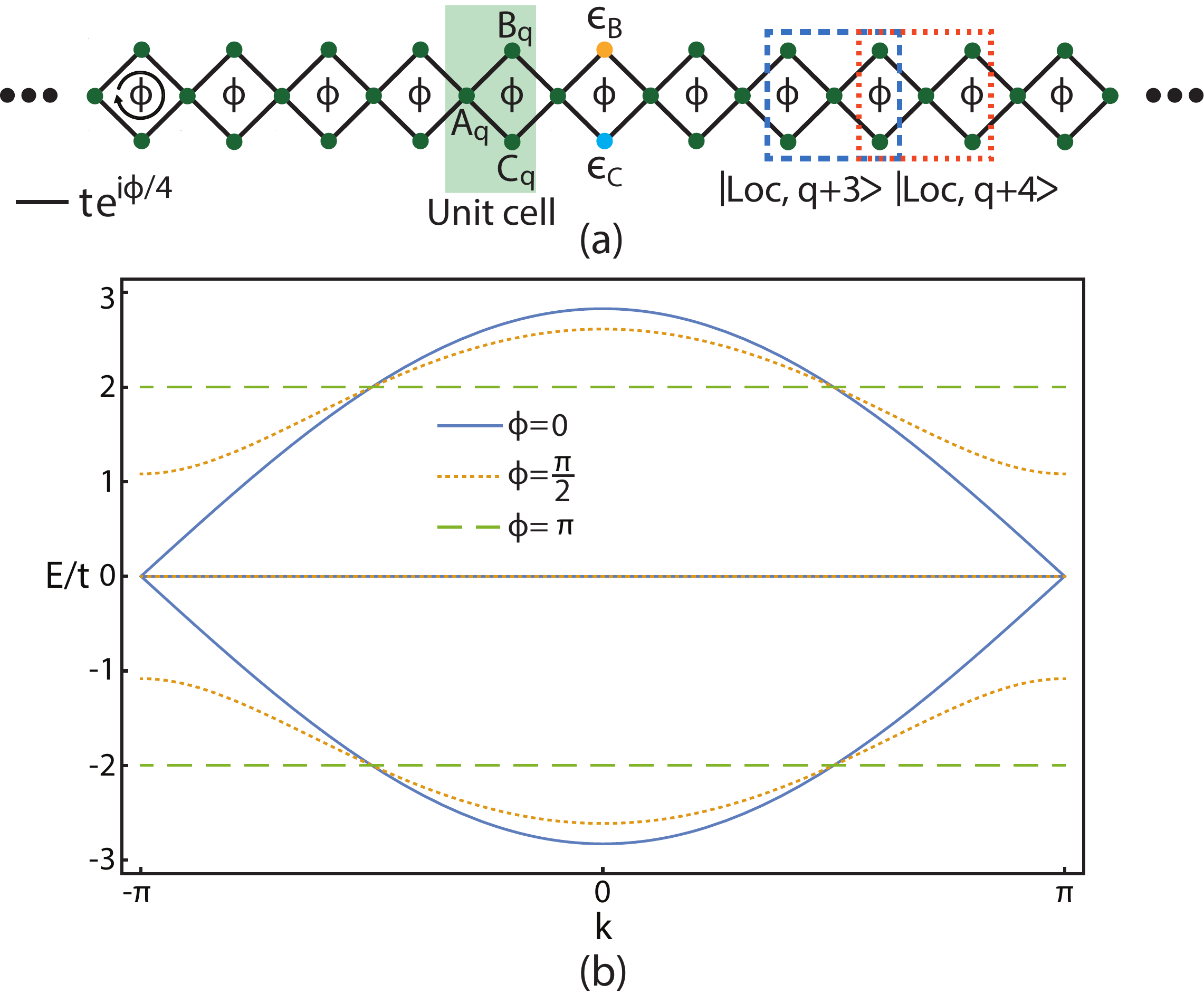} 
		\par\end{centering}
	\caption{\textbf{Diamond chain.} (a) Diamond chain with $\phi$ flux per plaquette, distributed uniformly by the four hopping terms of each plaquette, with positive phase accumulated in the clockwise direction. Green shaded region represents the $q$th unit cell with sites $A_q$, $B_q$ and $C_q$, while dashed blue and dotted orange regions represent the spatial width of adjacent CLSs $\ket{\text{Loc},q+3}$ and $	\ket{\text{Loc},q+4}$, respectively, which overlap in the middle $B$ and $C$ sites. Onsite impurity potentials $\epsilon_{_B}$ and $\epsilon_{_C}$ are placed at the orange $B_{q+1}$ and light blue $C_{q+1}$ sites, respectively. (b) Bulk energy spectrum, in units of $t$, of the diamond chain for three representative values of the $\phi$ flux per plaquette. The zero-energy flat band is common to all $\phi$ values.}
	\label{fig:chain}
\end{figure}

The specific system we will consider from now on will be that of Fig.~\ref{fig:chain}(a), consisting of a diamond chain with a $\phi=2\pi\frac{\Phi}{\Phi_0}$ reduced magnetic flux per plaquette, with $\Phi$ the magnetic flux and $\Phi_0$ the magnetic flux quantum.
Setting the lattice constant to unit, the bulk Hamiltonian of the clean system is given, in the $\{\ket{A(k)},\ket{B(k)},\ket{C(k)}\}$ basis, with $k$ the quasimomentum, by
\begin{eqnarray}
	H(k)&=&-t
	\begin{pmatrix}
		0&h
		\\
		h^\dagger&0
	\end{pmatrix},
\\
h^\dagger&=&
\begin{pmatrix}
	e^{i\frac{\phi}{4}}+e^{i(k-\frac{\phi}{4})}
	\\
	e^{-i\frac{\phi}{4}}+e^{i(k+\frac{\phi}{4})}
\end{pmatrix}
\end{eqnarray}
where $t$ is the magnitude of the hopping term.
Diagonalization of this Hamiltonian yields three energy bands, 
\begin{eqnarray}
	E_0&=&0,
	\\
	E_\pm(k)&=&\pm 2t\sqrt{1+\cos k\cos \frac{\phi}{2}},
\end{eqnarray}
where $E_\pm(k)$ are dispersive for all $\phi$ except for $\phi=\pi$, where they become flat [see Fig.~\ref{fig:chain}(b)].
An energy gap between the zero-energy FB $E_0$ and $E_\pm(k)$ opens at $k=\pi$ for $\phi\neq 0$, with a value of $E_{\text{gap}}=2t\sqrt{1-\cos\frac{\phi}{2}}$.

The eigenstates corresponding to $E_0$ can be written as CLSs, which for finite flux span over two plaquettes \cite{Liberto2019}, as shown in Fig.~\ref{fig:chain}(a). 
Consecutive normalized CLSs of the zero-energy FB are written, in the $\{\ket{B_j},\ket{C_j},\ket{B_{j+1}},\ket{C_{j+1}},\ket{B_{j+2}},\ket{C_{j+2}}\}$ basis, where $\ket{\zeta_l}$ stands for the state at site $\zeta=B,C$ of unit cell $l$ (note that $A$ sites are absent from this basis, as the CLSs have no weight on them), as
\begin{eqnarray}
	\ket{\text{Loc},j}&=&\frac{1}{2}\big(1,-e^{-i\frac{\phi}{2}},e^{-i\frac{\phi}{2}},-1,0,0\big)^T,
	\label{eq:locj}
	\\
	\ket{\text{Loc},j+1}&=&\frac{1}{2}\big(0,0,1,-e^{-i\frac{\phi}{2}},e^{-i\frac{\phi}{2}},-1\big)^T,
	\label{eq:locjplus1}
\end{eqnarray}
which overlap in the two middle sites [see Fig.~\ref{fig:chain}(a)].
By taking advantage of this overlap, it was already shown that projecting spatially modulated onsite interactions on the CLSs can induce topological behavior in a two-body system \cite{Pelegri2020}.
Here, we will focus on the single-particle case.
The elements of the overlap matrix within the non-orthogonal subspace of CLSs of the zero-energy FB, $S_{ij}=\braket{\text{Loc},i}{\text{Loc},j}$, are given, using (\ref{eq:locj})-(\ref{eq:locjplus1}), by
\begin{equation}
	\begin{cases}
		S_{jj}=1,
		\\
		S_{jj+1}=S_{j+1j}=\frac{1}{2}\cos\frac{\phi}{2},
		\\
		S_{ij}=0, \abs{i-j}>1,
	\end{cases}
\end{equation}
yielding a tridiagonal real and symmetric Toeplitz matrix,
\begin{equation}
	\mathbf{S}=\begin{pmatrix}
		1&a&
		\\
		a&1&a
		\\
		&a&1&\ddots
		\\
		&&\ddots&\ddots
		\\
		&&&&1&a&
		\\
		&&&&a&1&a
		\\
		&&&&&a&1
	\end{pmatrix},
\label{eq:smatrix}
\end{equation}
where the entries not shown are zeros and $a:=\frac{1}{2}\cos\frac{\phi}{2}$.
Notice how $a=0$ for $\phi=\pi$, such that $\mathbf{S}$ reduces to $\mathbf{I}$, that is, the CLSs form an orthogonal basis, even though adjacent CLSs still have spatial overlap \cite{Pelegri2020}.
The formula for the elements of the inverse of a tridiagonal matrix is known \cite{Fonseca2001}, and reduces, in our case of interest (that of $\mathbf{S}^{-1}$), to
\begin{equation}
	S_{nm}^{-1}=
	\begin{cases}
		(-1)^{n+m}\frac{1}{a}\frac{U_{n-1}(\frac{1}{2a}) U_{N-m}(\frac{1}{2a})}{U_{N}(\frac{1}{2a})},\ \ n\leq m,
		\\
		(-1)^{n+m}\frac{1}{a}\frac{U_{m-1}(\frac{1}{2a}) U_{N-n}(\frac{1}{2a})}{U_{N}(\frac{1}{2a})},\ \ n>m,
	\end{cases}
\label{eq:sinverse}
\end{equation}
where $U_n(x)$ is the Chebyshev polynomial of the second kind, defined as
\begin{equation}
	U_n(x)=\frac{\sinh\big[(n+1)\theta\big]}{\sinh\theta}, \ \ \abs{x=\cosh\theta}>1,
	\label{eq:chebyshev}
\end{equation}
whose argument, for the case of (\ref{eq:sinverse}), reads as
\begin{equation}
	x=\cosh\theta=\frac{1}{2a}=\sec\frac{\phi}{2},
	\label{eq:theta}
\end{equation}
Substituting (\ref{eq:chebyshev}) and (\ref{eq:theta}) in (\ref{eq:sinverse}) leads to
\begin{equation}
	S_{nm}^{-1}=(-1)^{n+m}2\coth\theta\frac{\sinh\big[n\theta\big]\sinh\big[(N-m+1)\theta\big]}{\sinh\big[(N+1)\theta\big]},
	\label{eq:sinvelements1}
\end{equation}
for $n\leq m$, and
\begin{equation}
	S_{nm}^{-1}=(-1)^{n+m}2\coth\theta\frac{\sinh\big[m\theta\big]\sinh\big[(N-n+1)\theta\big]}{\sinh\big[(N+1)\theta\big]},
	\label{eq:sinvelements2}
\end{equation}
for $n>m$, such that $S_{ij}^{-1}=S_{ji}^{-1}$.
Note that the tridiagonal and repeating structure of $\mathbf{S}$ in (\ref{eq:smatrix}) is formally equivalent to that of an open linear chain with uniform couplings and a constant energy shift.
The analytical solution of these types of linear 1D models has been recently derived  \cite{Marques2020} and, therefore, one could alternatively find $\mathbf{S}^{-1}$ through the known eigenvalues and eigenvectors of $\mathbf{S}$.

\subsection*{Impurity states}
\label{sec:singleimp}
In this section, we will study the case, depicted in Fig.~\ref{fig:chain}(a), where impurities affect the local potentials at the top and bottom sites of a single plaquette in the middle of an open diamond chain with $N+1$ unit cells (such that one can construct $N$ zero-energy CLSs).
The corresponding impurity operator has the general form
\begin{equation}
	\hat{V}=\epsilon_{_B}\ket{B_{q+1}}\bra{B_{q+1}}+\epsilon_{_C}\ket{C_{q+1}}\bra{C_{q+1}},
	\label{eq:impoperator}
\end{equation}
where $\epsilon_{_B}$ and $\epsilon_{_C}$ are local impurity potentials, left as free parameters, and $q=\frac{N}{2}$, with $N$ assumed even.
We consider a finite flux per plaquette, $\phi\neq 0$, to have a finite $E_{\text{gap}}$ between the flat and dispersive bands, and also that $\epsilon_{_B},\epsilon_{_C}\ll E_{\text{gap}}$, such that higher-order terms mediated by the dispersive bands can be neglected.
Through (\ref{eq:projoperator}), the projected impurity operator onto the non-orthogonal FB subspace is written as
\begin{eqnarray}
	\hat{V}_{_{FB}}&=&\hat{P}^\dagger\hat{V}\hat{P} \nonumber
	\\
	&=&\sum\limits_{n,m=q}^{q+1}\ket{\text{Loc},n^*}V_{nm}\bra{\text{Loc},m^*},
	\label{eq:projfb}
\end{eqnarray}
where $V_{nm}=\bra{\text{Loc},n}\hat{V}\ket{\text{Loc},m}$ and the sums run only from $q$ to $q+1$ since, from the form of the CLSs in (\ref{eq:locj})-(\ref{eq:locjplus1}) and of $\hat{V}$ in (\ref{eq:impoperator}), only the CLSs with these indices are affected by the impurity potentials.
One can readily evaluate all the relevant $V_{nm}$ elements, which reduce to the $\{V_{qq},V_{qq+1},V_{q+1q},V_{q+1q+1}\}$ set, to obtain
\begin{eqnarray}
	\hat{V}_{_{FB}}&=&v\sum\limits_{n=q}^{q+1}\ket{\text{Loc},n^*}\bra{\text{Loc},n^*}+ \nonumber
	\\
	&+&\Big[w\ket{\text{Loc},q+1^*}\bra{\text{Loc},q^*}+\text{H.c.}\Big],
	\label{eq:impoperatorfb}
\end{eqnarray}
with $v=\frac{\epsilon_{_B}+\epsilon_{_C}}{4}$ acting as a local potential on the dual CLSs at $q$ and $q+1$, and $w=\big(\epsilon_{_B}e^{-i\frac{\phi}{2}}+\epsilon_{_C}e^{i\frac{\phi}{2}}\big)/4$ a complex hopping term between the states.

At this point the reader might suspect of a circular reasoning being employed here.
After all, the impurities placed at the midchain plaquette in (\ref{eq:impoperator}) only affect two adjacent CLSs, with indices $q$ and $q+1$, which are part of the non-orthogonal set of CLSs.
When we translate the system into the dual basis, we arrive at the effective dimer model of (\ref{eq:impoperatorfb}) which, again, only involves the two dual CLSs $q$ and $q+1$ that, in turn, are themselves also a part of the non-orthogonal set of dual CLSs.
It appears that one can go back and forth between the original and dual basis indefinitely, without moving the problem forward.

What, then, is to be gained by this procedure?
We can always restrict ourselves, in either basis, to the two states set $\{\ket{\text{Loc},q^{(*)}},\ket{\text{Loc},q+1^{(*)}}\}$, and assume that the rest of the localized states $\{\ket{\text{Loc},l^{(*)}}\}$, where $l\neq q,q+1$, are orthogonalized first with respect to the former two, and then between themselves.
However, and crucially, orthogonalization of this residual set, in the original basis, produces states of varying spatial extension (without losing their localized nature \cite{Lopes2014}), \textit{i.e.}, that spread to more than two consecutive plaquettes.
Due to this, several states in the orthogonalized set $\{\ket{\text{Loc},l}\}$ will have a finite weight at the $q+1$ plaquette where the impurities are located, such that, from (\ref{eq:impoperator}), we will have $V_{nm}\neq 0$ at least for some $n,m\neq q,q+1$, and the effective model is not restricted to two states in the original basis. 
On the other hand, the impurity operator of the dual basis, given in (\ref{eq:impoperatorfb}), is not spatially defined, but instead directly couples the $q$ and $q+1$ dual CLSs.
As such, regardless of the particular spatial profile of the dual states in the orthogonalized $\{\ket{\text{Loc},l^*}\}$ set, all effective impurity elements involving these states, computed from (\ref{eq:impoperatorfb}), will vanish, given that they are assumed orthogonal to the $\{\ket{\text{Loc},q^{*}},\ket{\text{Loc},q+1^{*}}\}$ states appearing in $\hat{V}_{_{FB}}$.
Therefore, working in the dual basis greatly simplifies the problem, now reduced to the diagonalization of a dimer model defined in a two-state dual basis.
\begin{figure*}[ht]
	\begin{centering}
		\includegraphics[width=0.95 \textwidth]{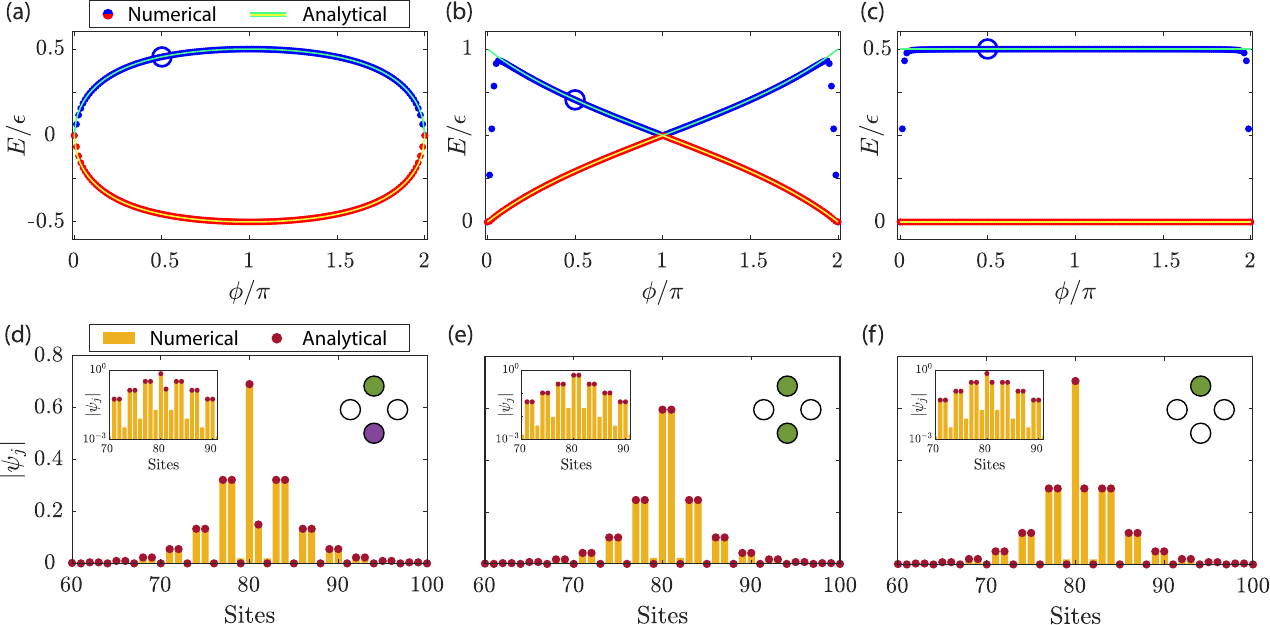} 
		\par\end{centering}
	\caption{\textbf{Impurity states.} Analytical energy curves (green and yellow curves) obtained from diagonalization of the effective two-state system in (\ref{eq:effimpmatrix}), as a function of the magnetic flux $\phi$ for (a) opposite impurities $\epsilon_{_B}=-\epsilon_{_C}=\epsilon$, (b) equal impurities $\epsilon_{_B}=\epsilon_{_C}=\epsilon$, and (c) a single impurity, $\epsilon_{_B}=\epsilon$ and $\epsilon_{_C}=0$, with $\epsilon=0.1t$.
		Blue and red dots correspond to the numerically obtained energies of the two states of the FB that are affected by the impurity, for an open diamond chain with 53 plaquettes pierced by a $\phi$ flux, with the impurities placed at the $B$ and $C$ sites of the midchain plaquette in (a) and (b), and at its $B$ site in (c).
		(d)-(f) Probability amplitudes of the eigenstate marked by a blue circle in the corresponding cases above. 
		Only the central part of the chain is shown.
		Golden bars (red dots) correspond to numerical (analytical) results.
		The left insets plot the same results in a semi-logarithmic scale and for a smaller portion of the chain.
		Sites without red dots are $A$ sites where the CLSs forming the analytical subspace have zero weight.
		The right insets depict the impurities at the midchain plaquette, with green (purple) color indicating an $\epsilon$ ($-\epsilon$) value for the impurity at that site.
	}
	\label{fig:diamond}
\end{figure*}

Since the two-state dual basis $\{\ket{\text{Loc},q^*},\ket{\text{Loc},q+1^*}\}$ is itself non-orthogonal, we first need to orhogonalize it through the L\"owdin transformation \cite{Loewdin1956,Leschber1989}, as outlined in the ``Symmetric orthonormalization'' subsection of the Results.
The first step requires us to find the form of $\mathbf{S}_*^{-\frac{1}{2}}$, where $\mathbf{S}_*$ is the overlap matrix written in this basis.
Using (\ref{eq:directdualtransf}), it is straightforward to show that the elements of $\mathbf{S}_*$ can be expressed as
\begin{eqnarray}
	S_{*,nm}&=&\braket{\text{Loc},n^*}{\text{Loc},m^*} \nonumber
	\\
	&=&\sum\limits_{i,j}S_{ni}^{-1}\braket{\text{Loc},i}{\text{Loc},j}S_{jm}^{-1} \nonumber
	\\
	&=&\sum\limits_j\Big(\sum\limits_i S_{ni}^{-1}S_{ij}\Big)S_{jm}^{-1} \nonumber
	\\
	&=&\sum\limits_j \delta_{nj}S_{jm}^{-1} \nonumber
	\\
	&=&S_{nm}^{-1},
\end{eqnarray}
where (\ref{eq:deltakron}) was used in the fourth line.
Using $q=\frac{N}{2}$ and assuming the thermodynamic limit ($N\to \infty$), such that $\sinh \big[x(N)\big]\to \frac{e^{x(N)}}{2}$ in (\ref{eq:sinvelements1})-(\ref{eq:sinvelements2}), we obtain the following expressions for some elements of the inverse overlap matrix,
\begin{equation}
	\begin{cases}
		S_{q+\Delta q+\Delta}^{-1}=\coth\theta
		\\
		S_{qq+\Delta}^{-1}=(-1)^{\Delta}\coth\theta e^{-\Delta\theta}
	\end{cases},\ \ \Delta=0,1,2,\dots,
\label{eq:sinvelementsq}
\end{equation}
with $S_{qq+\Delta}^{-1}=S_{q+\Delta q}^{-1}$.
With the formulas of (\ref{eq:sinvelementsq}), we find the form of $\mathbf{S}_*^{-\frac{1}{2}}$,
\begin{eqnarray}
	\mathbf{S}_*^{-\frac{1}{2}}&=&
	\begin{pmatrix}
		S_{qq}^{-1}&S_{qq+1}^{-1}
		\\
		S_{qq+1}^{-1}&S_{q+1q+1}^{-1}
	\end{pmatrix}^{-\frac{1}{2}} \nonumber
\\
&=&
	\begin{pmatrix}
	\coth\theta&-\coth\theta e^{-\theta}
	\\
	-\coth\theta e^{-\theta}&\coth\theta
\end{pmatrix}^{-\frac{1}{2}}  \nonumber
\\
&=&
\frac{1}{2\sqrt{\coth \theta}}
	\begin{pmatrix}
	\alpha_-+\alpha_+&\alpha_--\alpha_+
	\\
	\alpha_--\alpha_+&\alpha_-+\alpha_+
\end{pmatrix},
\label{eq:sdualm12}
\end{eqnarray}
with $\alpha_\pm=\frac{1}{\sqrt{1\pm e^{-\theta}}}$.
Combining (\ref{eq:lowdin}) and (\ref{eq:sdualm12}), we write the orthogonal two-state basis $\{\ket{\tilde{\varphi}_q},\ket{\tilde{\varphi}_{q+1}}\}$ as
\begin{eqnarray}
	\ket{\tilde{\varphi}_q}&=&\sum\limits_{j=q}^{q+1}S^{-\frac{1}{2}}_{*,qj}\ket{\text{Loc},j^*}\nonumber
 \\
 &=&\sum\limits_{j=q}^{q+1}\sum\limits_{l}S^{-\frac{1}{2}}_{*,qj}S^{-1}_{jl}\ket{\text{Loc},l},
	\label{eq:tildephiq}
	\\
	\ket{\tilde{\varphi}_{q+1}}&=&\sum\limits_{j=q}^{q+1}S^{-\frac{1}{2}}_{*,q+1j}\ket{\text{Loc},j^*}\nonumber
 \\
 &=&\sum\limits_{j=q}^{q+1}\sum\limits_{l}S^{-\frac{1}{2}}_{*,q+1j}S^{-1}_{jl}\ket{\text{Loc},l}.
	\label{eq:tildephiqplus1}
\end{eqnarray}
In turn, the impurity matrix is constructed from (\ref{eq:impoperatorfb}) as
\begin{equation}
	\mathbf{V}_{_{FB}}=
		\begin{pmatrix}
		V_{*,qq}&V_{*,qq+1}
		\\
		V_{*,q+1q}&V_{*,q+1q+1}
	\end{pmatrix},
\label{eq:impmatrix}
\end{equation}
with $V_{*,nm}=\bra{\text{Loc},n^*}\hat{V}_{_{FB}}\ket{\text{Loc},m^*}$ and
\begin{widetext}
\begin{eqnarray}
	V_{*,qq}&=&V_{*,q+1q+1}=v\big[(S_{qq}^{-1})^2+(S_{qq+1}^{-1})^2\big]+(w+\bar{w})S_{qq}^{-1}S_{qq+1}^{-1},
	\\
	V_{*,qq+1}&=&V_{*,q+1q}^*=\bar{w}(S_{qq}^{-1})^2+w(S_{qq+1}^{-1})^2+2vS_{qq}^{-1}S_{qq+1}^{-1},
	\label{eq:v*qqplus1}
\end{eqnarray}
\end{widetext}
where $\bar{w}$ is the complex conjugate of $w$.
According to (\ref{eq:effhamilt}), the effective impurity matrix is computed from (\ref{eq:sinvelementsq}), (\ref{eq:sdualm12}) and (\ref{eq:impmatrix}) and yields
\begin{widetext}
\begin{eqnarray}
	\mathbf{\tilde{V}}_{_{FB}}&=&\mathbf{S}_*^{-\frac{1}{2}}\mathbf{V}_{_{FB}}\mathbf{S}_*^{-\frac{1}{2}} \nonumber
	\\
	&=&
	\begin{pmatrix}
		\frac{\epsilon_{_B}+\epsilon_{_C}}{4}
		&
		\frac{\epsilon_{_B}+\epsilon_{_C}}{4}e^{-\theta} +i\frac{\epsilon_{_B}-\epsilon_{_C}}{4}\sqrt{1-e^{-2\theta}}
		\\					
		\frac{\epsilon_{_B}+\epsilon_{_C}}{4}e^{-\theta}-i\frac{\epsilon_{_B}-\epsilon_{_C}}{4}\sqrt{1-e^{-2\theta}}
		&
		\frac{\epsilon_{_B}+\epsilon_{_C}}{4}
	\end{pmatrix}
    \label{eq:effimpmatrix}.
\end{eqnarray}
\end{widetext}
Diagonalization of this matrix yields, through (\ref{eq:theta}), the flux-dependent energies $E_\pm=E_\pm(\phi)$ (which will be referred hereafter as the ``energy bands'', with the flux taking the place of a synthetic momentum)  of the two states $\ket{\tilde{\varphi}_\pm}=\ket{\tilde{\varphi}_\pm(\phi)}$ within the zero-energy FB that are affected by the impurities at the $q+1$ plaquette, which are determined as
\begin{equation}
	\begin{pmatrix}
		\ket{\tilde{\varphi}_+}
		\\
		\ket{\tilde{\varphi}_-}
	\end{pmatrix}
=\mathbf{U}
	\begin{pmatrix}
	\ket{\tilde{\varphi}_q}
	\\
	\ket{\tilde{\varphi}_{q+1}}
\end{pmatrix},
\end{equation}
where $\mathbf{U}$ is the unitary matrix with eigenvectors as its rows, with elements $[U]_{ij}\equiv U_{ij}$, where $i,j=q,q+1$.
Using (\ref{eq:tildephiq})-(\ref{eq:tildephiqplus1}), we can express $\ket{\tilde{\varphi}_\pm}$ in terms of the original CLSs,
\begin{eqnarray}
	\ket{\tilde{\varphi}_+}&=&\sum\limits_{j=q}^{q+1}\sum\limits_{l}\Big[U_{qq}S^{-\frac{1}{2}}_{*,qj}+U_{qq+1}S^{-\frac{1}{2}}_{*,q+1j}\Big]S^{-1}_{jl}\ket{\text{Loc},l},
	\label{eq:tildephiplus}
	\\
	\ket{\tilde{\varphi}_-}&=&\sum\limits_{j=q}^{q+1}\sum\limits_{l}\Big[U_{q+1q}S^{-\frac{1}{2}}_{*,qj}+U_{q+1q+1}S^{-\frac{1}{2}}_{*,q+1j}\Big]S^{-1}_{jl}\ket{\text{Loc},l}.
	\label{eq:tildephiminus}
\end{eqnarray}

Let us address separately three representative choices for the $\epsilon_{_B},\epsilon_{_C}$ local impurities shown in the right insets of Fig.~\ref{fig:diamond}(d)-(f) and analyze their solutions.

\subsubsection*{Opposite impurities}

\begin{figure*}[ht]
	\begin{centering}
		\includegraphics[width=0.98 \textwidth]{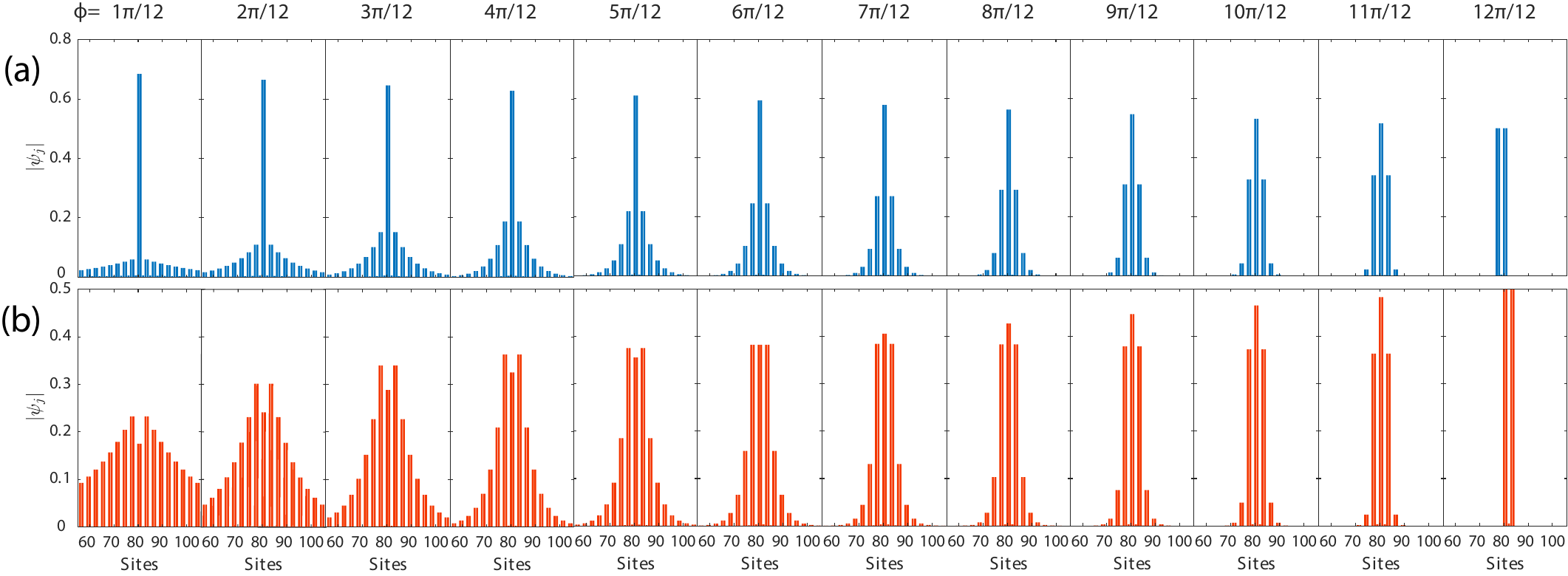} 
		\par\end{centering}
	\caption{\textbf{Evolution of impurity states with flux.} Snapshots of the amplitudes of the (a) higher and (b) lower impurity state at different flux values, indicated on top, for a partial section of an open diamond chain with 53 complete plaquettes, with equal impurities $\epsilon=0.1t$ in the middle one. The energy of both states at each flux value is given in Fig.~\ref{fig:diamond}(b).}
	\label{fig:amplitudes}
\end{figure*}
For opposite impurities, $\epsilon_{_B}=-\epsilon_{_C}=\epsilon$, where the impurity strength $\epsilon$ is given in units of the hopping parameter $t$, the effective impurity matrix in (\ref{eq:effimpmatrix}) reduces to
\begin{equation}
	\mathbf{\tilde{V}}_{_{FB}}=-\frac{\epsilon}{2}\sqrt{1-e^{-2\theta}}\sigma_y,
\end{equation}
with $\sigma_\mu$ ($\mu=x,y,z$) the $\mu$ Pauli matrix.
The respective eigenvalues and eigenvectors read as
\begin{eqnarray}
	E_\pm&=&\pm\frac{\epsilon}{2}\sqrt{1-e^{-2\theta}},
	\label{eq:enersym}
	\\
	\begin{pmatrix}
	\ket{\tilde{\varphi}_+}
	\\
	\ket{\tilde{\varphi}_-}
\end{pmatrix}
&=&\frac{1}{\sqrt{2}}
\begin{pmatrix}
	i&1
	\\
	-i&1
\end{pmatrix}
\begin{pmatrix}
	\ket{\tilde{\varphi}_q}
	\\
	\ket{\tilde{\varphi}_{q+1}}
\end{pmatrix}.	
\label{eq:eigssymimps}
\end{eqnarray}
In Fig.~\ref{fig:diamond}(a), we compare the analytical energies of (\ref{eq:enersym}) with numerical results as a function of the magnetic flux, for a chain with 53 plaquettes and opposite local impurities at the top and bottom sites of the midchain plaquette [depicted in the right inset of Fig.~\ref{fig:diamond}(d)].
Two states with opposite energies are lifted from the flat band for a finite  magnetic flux per plaquette.
Even though the perturbation is already quite large, $\epsilon=0.1t$, that is, away for all $\phi$ from the $\epsilon\ll E_{\text{gap}}$ condition assumed in the projection of the impurity operator onto the flat band in (\ref{eq:projfb}), there is a nearly perfect match between analytical and numerical curves for the entire spectrum.

In Fig.~\ref{fig:diamond}(d), we plot the spatial profile of the eigenstate corresponding to the encircled positive energy at $\phi=\frac{\pi}{2}$ in Fig.~\ref{fig:diamond}(a).
The analytical amplitude at each site, depicted as red dots, was computed by applying the general expression in (\ref{eq:tildephiplus}), with the $\mathbf{U}$ matrix for the opposite impurities case given in (\ref{eq:eigssymimps}), and agrees very well with the numerical one, represented by the golden bar at the corresponding site.
The inset shows the same results but in a semi-logarithmic scale.
Sites without red dots correspond to A sites where the localized states have zero weight, according to (\ref{eq:locj})-(\ref{eq:locjplus1}).
The negligible numerical weight at these sites, which is at least an order of magnitude smaller than the adjacent B and C sites with highest weight, comes from higher-order terms that couple the perturbed states with the top and bottom dispersive bands. 
Therefore, as the impurity value is decreased to approach the $\epsilon\ll E_{\text{gap}}$ limit, the weight at the A sublattice decreases even further, relative to the weight at the BC sublattice.

\subsubsection*{Equal impurities}

For equal impurities, $\epsilon_{_B}=\epsilon_{_C}=\epsilon$, the effective impurity matrix in (\ref{eq:effimpmatrix}) reduces to
\begin{equation}
	\mathbf{\tilde{V}}_{_{FB}}=\frac{\epsilon}{2}\big(\sigma_0 + e^{-\theta}\sigma_x\big),
\end{equation}
where $\sigma_0$ is the $2\times 2$ identity matrix.
The respective eigenvalues and eigenvectors read as
\begin{eqnarray}
	E_\pm&=&\frac{\epsilon}{2}\big(1\pm e^{-\theta}\big),
	\label{eq:enereq}
	\\
	\begin{pmatrix}
		\ket{\tilde{\varphi}_+}
		\\
		\ket{\tilde{\varphi}_-}
	\end{pmatrix}
	&=&\frac{1}{\sqrt{2}}
	\begin{pmatrix}
		1&1
		\\
		1&-1
	\end{pmatrix}
	\begin{pmatrix}
		\ket{\tilde{\varphi}_q}
		\\
		\ket{\tilde{\varphi}_{q+1}}
	\end{pmatrix}.	
	\label{eq:eigseqimps}
\end{eqnarray}

In Fig.~\ref{fig:diamond}(b), we compare the analytical energies of (\ref{eq:enereq}) with numerical results as a function of the magnetic flux for the impurity distribution depicted in the right inset of Fig.~\ref{fig:diamond}(e).
Two positive energy states are lifted from the FB for a finite  magnetic flux per plaquette, and become degenerate with $E=0.5\epsilon$ at $\phi=\pi$.
Since $E_{\text{gap}}\to 0$ when $\phi\to 0\ (\text{mod}\ 2\pi)$, near the edges of the spectrum higher-order couplings to the dispersive bands can no longer be safely neglected, and the analytical treatment breaks down in that region.
However, this region is still very small for $\epsilon=0.1t$, and shrinks even further with decreasing $\epsilon$.
The spatial amplitudes of the eigenstate with the energy marked by the blue circle in Fig.~\ref{fig:diamond}(b) are depicted in Fig.~\ref{fig:diamond}(e).
As for the case of opposite impurities above, a very good agreement is found between the numerical and analytical results.

\begin{figure}[ht]
		\begin{centering}
			\includegraphics[width=0.925 \columnwidth]{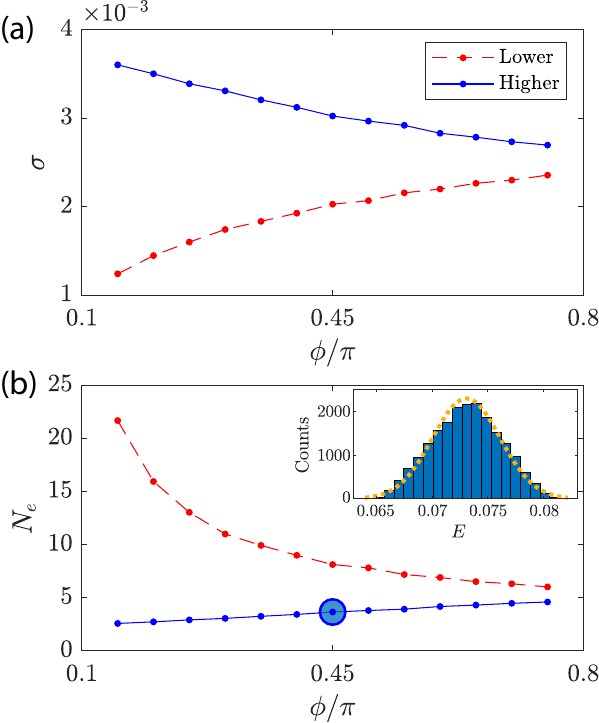} 
			\par\end{centering}
		\caption{\textbf{Disorder effects in the impurity subspace.} (a) Standard deviation, as a function of the flux, of the energy of both impurity states for an open diamond chain with 101 plaquettes, with equal impurities $\epsilon=0.1t$ in the middle one, taken from 20000 realizations of diagonal disorder at each point, with disorder strength $\delta=0.02t$.
			(b) Effective number of sites involved in the disorder averaging effect, computed from the corresponding standard deviation values in (a) in accordance with the CLT.
			The inset shows the energy histogram for the encircled higher-energy state at $\phi=0.45\pi$, with the dashed orange curve corresponding to the Gaussian fit from which the standard deviation is extracted.}
		\label{fig:sigmas}
	\end{figure}

Having established numerically the validity of the analytical solutions considerably away from the $\epsilon\ll E_{\text{gap}}$ limit assumed in their derivation (the impurity strength is $\epsilon=0.1t$ in Fig.~\ref{fig:diamond}, and even higher values still yield good agreement with numerical results), one can study the effect of small disorder on the impurity states.
More specifically, we will consider diagonal disorder on an open diamond chain with 101 plaquettes threaded by a $\phi$ flux, by setting the onsite potential at each site to a random value taken from a uniform distribution in the interval $[-\frac{\delta}{2},\frac{\delta}{2}]$, with $\delta=0.02t$ the disorder strength, with a standard deviation given by $\sigma_{\text{ud}}=\frac{\delta}{\sqrt{12}}\approx 5.8\times 10^{-3}$.
For the sites at the midchain plaquette hosting the impurities, the disorder is added on top of $\epsilon=0.1t$.	
We perform 20000 disorder realizations for each flux value and track the energies of the impurity states, from where we construct their respective histograms, as exemplified for the higher one at $\phi=\frac{9\pi}{20}$ in the inset of Fig.~\ref{fig:sigmas}(b).
The underlying uniform distribution of the disorder generates a normal distribution for the energies of the impurity states, whose mean approaches their disorder-free values given in Fig.~\ref{fig:diamond}(b) with increasing disorder realizations.
This result is consistent with the central limit theorem (CLT), \textit{i.e.}, for each realization the impurity states effectively feel an averaged disorder related to their localization length, that is, to their spatial extension, which is controlled by the flux, as illustrated in Fig.~\ref{fig:amplitudes}.
	
In Fig.~\ref{fig:sigmas}(a), we plot the standard deviation of the Gaussian fit of the energy distributions of both impurity states as a function of the flux .
The flux was taken in $0.05\pi$ increments  and restricted to the $[0.15\pi,0.75\pi]$ interval, since near $\phi=0,\pi$ the energy separation of the states is of the order of $\delta$ [see numerical results in Fig.~\ref{fig:diamond}(b) in these regions], which leads to a reordering of the states and to an overlap of their corresponding energy distributions, preventing their independent analysis.
According to the CLT, the standard deviation of the normal distribution is related to the sample size, which in our case corresponds to the effective number of sites $N_e$ over which the disorder is averaged for each impurity state, through $\sigma=\frac{\sigma_{\text{ud}}}{\sqrt{N_e}}$, with $\sigma_{\text{ud}}$ given above.
With this formula, and from the $\sigma$ values given in Fig.~\ref{fig:sigmas}(a), we plot in Fig.~\ref{fig:sigmas}(b) the flux dependence of $N_e$ for both impurity states.
It can be seen that they follow opposite tendencies with increasing flux: while $N_e$ decreases exponentially for the lower state, it increases monotonously for the higher state.

The evolution of the spatial profile of the impurity states with flux, shown in Fig.~\ref{fig:amplitudes}, can be used to explain their different response to disorder. 
The lower state becomes globally more and more localized until it becomes a CLS at $\phi=\pi$, which agrees with its decreasing $N_e$ in Fig.~\ref{fig:sigmas}(b),
since the disorder is being averaged over fewer sites, therefore generating a higher effective onsite disorder on the lower state.
The higher state, on the other hand, has dominant contributions from the impurity sites at lower fluxes (see central peaks in the top panel of Fig.~\ref{fig:amplitudes} with growing weight for decreasing flux).
The weights at these sites are in an anti-symmetric configuration that becomes the full eigenstate (a CLS) at $\phi=0$, while they are in a symmetric configuration for the lower state.
At zero flux, applying the same rotation to the $B$ and $C$ sites at each plaquette has been shown to decouple the anti-symmetric state, while the symmetric one connects with adjacent $A$ sites to form a uniform chain \cite{Leykam2017b,Nicolau2023b}.
As flux is increased, the anti-symmetric configuration of impurity sites couples to the uniform chain with growing strength \cite{Lopes2011} and, as with the lower state, the higher state develops a spatially extended profile that becomes increasingly localized (see the evolution of the components at sites other than the central impurity ones in the top panel of Fig.~\ref{fig:amplitudes}).
Conversely, the peaks at the impurity sites, which are the main contributors to the number of effective sites $N_e$ of the averaged disorder, continuously decrease with increasing flux, that is, spread their weight to the other sites.
This effect, which tends to increase $N_e$, overpowers the localization effect of the other components, which tends to decrease $N_e$ (as for the lower state), leading to the net increase of $N_e$ with flux seen in Fig.~\ref{fig:sigmas}(b).

While the normal distribution of the sample means expected from CLT is only exact, strictly speaking, in the limit of infinite sample size, $N_e\to\infty$, we have checked numerically that a Gaussian fit is in very good agreement with the energy histogram of all points in Fig.~\ref{fig:sigmas}(b), including for the higher state, as exemplified in the inset, where the effective number of sites responsible for the averaged disorder is small, ranging from $N_e\approx 2.5$ to $N_e\approx 5$.

The averaging of diagonal disorder on the impurity states, conjectured to be the mechanism behind the enhanced robustness to disorder displayed by certain domain wall states with similar localization properties \cite{Munoz2018}, is quantitatively confirmed here by the numerical results of Fig.~\ref{fig:sigmas}.
For the equal impurities case studied in this section, this enhanced robustness to disorder is greater for the lower impurity state at low flux values where, according to Fig.~\ref{fig:sigmas}(b), $N_e$ is maximized.
This is demonstrated in a follow up paper, where the response to disorder of effective systems of coupled impurity states is analyzed in greater detail \cite{Viedma2024b}.

As for the finite energy impurity states studied in the ``Opposite impurities'' subsection above and  in the ``Single impurity'' subsection below, we have checked numerically that this averaging effect is less pronounced.
Under the same kind of diagonal disorder and for different flux values, the energy histogram of these other impurity states yields in general a cross between a Gaussian and a uniform distribution, signaling a number of effective sites involved in the averaging effect close to one, such that the CLT cannot be used to determine this number.
This can be explained from the dominant weight that these states always have at one of the impurity sites [see the central peak in Figs.~\ref{fig:diamond}(d) and (f)], which is where the perturbative effects of disorder are felt the most by these states.
Therefore, the CLT only provides an accurate description of the averaging effect of diagonal disorder for the equal impurities case.

\subsubsection*{Single impurity}

\begin{figure}[ht]
	\begin{centering}
		\includegraphics[width=0.95 \columnwidth]{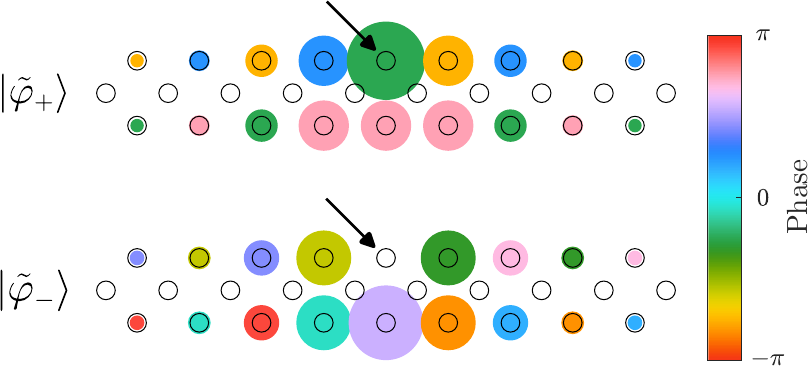} 
		\par\end{centering}
	\caption{\textbf{Midchain profile of impurity states.} Partial profile of the $\ket{\tilde{\varphi}_+}$ and $\ket{\tilde{\varphi}_-}$ eigenstates at $\phi=\frac{\pi}{2}$, where the radius of each circle represents the amplitude of the wavefunction at the respective site and the color represents its phase, coded by the color bar at the right.
		The arrow indicates the $B_{q+1}$ site where the $\epsilon_{_B}=\epsilon=0.1t$ impurity is placed, and where $\ket{\tilde{\varphi}_+}$ ($\ket{\tilde{\varphi}_-}$) has maximum (zero) amplitude.}
	\label{fig:singleimpstates}
\end{figure}

Finally, let us address the case where a single impurity, $\epsilon_{_B}=\epsilon$, is placed at the top site of the $q+1$ plaquette.
The effective impurity matrix in (\ref{eq:effimpmatrix}) reduces now to
\begin{eqnarray}
	\mathbf{\tilde{V}}_{_{FB}}&=&\frac{\epsilon}{4}\big(\sigma_0 + e^{-\theta}\sigma_x -\sqrt{1-e^{-2\theta}}\sigma_y\big).
	\label{eq:hamiltsingleimp}
\end{eqnarray}
The respective eigenvalues and eigenvectors read as
\begin{widetext}
\begin{eqnarray}
	E_\pm&=&\frac{\epsilon}{4}(1\pm 1),
	\label{eq:enersingle}
	\\
	\begin{pmatrix}
		\ket{\tilde{\varphi}_+}
		\\
		\ket{\tilde{\varphi}_-}
	\end{pmatrix}
	&=&\frac{1}{\sqrt{2}}
	\begin{pmatrix}
		1&e^{-i\vartheta}
		\\
        1&-e^{-i\vartheta}
	\end{pmatrix}
	\begin{pmatrix}
		\ket{\tilde{\varphi}_q}
		\\
		\ket{\tilde{\varphi}_{q+1}}
	\end{pmatrix},	
	\label{eq:eigssingleimp}\
	\\
	\cos\vartheta&:=&e^{-\theta}=\text{Exp}\Big[-\cosh^{-1}\big[\sec\frac{\phi}{2}\big]\Big],
	\label{eq:vartheta}
\end{eqnarray}
\end{widetext}
where the formula for $\theta$ given in (\ref{eq:theta}) was used in the last line.
Both energy bands are flat (in $\phi$) and, in particular, we have $E_-=0$, as shown in Fig.~\ref{fig:diamond}(c), meaning that, when a single impurity is present, the system can always rotate the subspace of dual localized states in order to generate a state, namely $\ket{\tilde{\varphi}_-}$ in (\ref{eq:eigssingleimp}), with zero weight at the impurity site, $\braket{B_{q+1}}{\tilde{\varphi}_-}=0$, such that it remains degenerate with the other FB states.
This is illustrated in Fig.~\ref{fig:singleimpstates}, where a partial profile of the $\ket{\tilde{\varphi}_+}$ and $\ket{\tilde{\varphi}_-}$ eigenstates, computed through (\ref{eq:eigssingleimp}) for $\phi=\frac{\pi}{2}$, is depicted [the complete profile of $\ket{\tilde{\varphi}_+}$ is shown in Fig.~\ref{fig:diamond}(f), which agrees very well with numerics].
It can be seen that the weight of the state at the impurity site is maximum for $\ket{\tilde{\varphi}_+}$ and zero for $\ket{\tilde{\varphi}_-}$.

We turn now to the topological characterization of the single-impurity case.
But first let us summarize the usual approach to the topological characterization of gapped 1D two-band models.
The relevant topological invariant for characterizing the energy gap is the Zak phase of the lower band, which can be expressed as
\begin{equation}
	\gamma=i\oint dR\bra{\tilde{\varphi}(R)_-}\frac{d}{dR}\ket{\tilde{\varphi}(R)_-},
	\label{eq:zak}
\end{equation}
where the integral traces a closed path along the dimensional parameter $R$.
For the textbook Su-Schrieffer-Heeger (SSH) two-band model \cite{Su1979}, consisting of a chain with staggered hopping terms $t_1$ and $t_2$, with the former (latter) corresponding to the intracell (intercell) hopping parameter \cite{Asboth2016}, the lowest energy bulk eigenstate has the same form of $\ket{\tilde{\varphi}_-}$ in (\ref{eq:eigssingleimp}) when written in its site state basis, $\ket{\tilde{\varphi}_-(k)}=1/\sqrt{2}(1,-e^{-i\vartheta(k)})^T$, where $k$ is the momentum and $\cos \vartheta(k)=\frac{t_1}{t_2\sin k}+\cot k$ \cite{Delplace2011,Marques2020}.
Inserting $\ket{\tilde{\varphi}_-(k)}$ in (\ref{eq:zak}), where $R=k$ and the integral goes over the Brillouin zone (BZ), we get
\begin{equation}
	\gamma=\frac{1}{2}\int_{\vartheta(0)}^{\vartheta(2\pi)}d\vartheta(k)=\frac{\Delta\vartheta}{2},
	\label{eq:zakfinal}
\end{equation}
with $\Delta\vartheta=\vartheta(2\pi)-\vartheta(0)$ yielding the total change in the $\vartheta(k)$ phase around the BZ.
\begin{figure}[ht]
	\begin{centering}
		\includegraphics[width=0.9 \columnwidth]{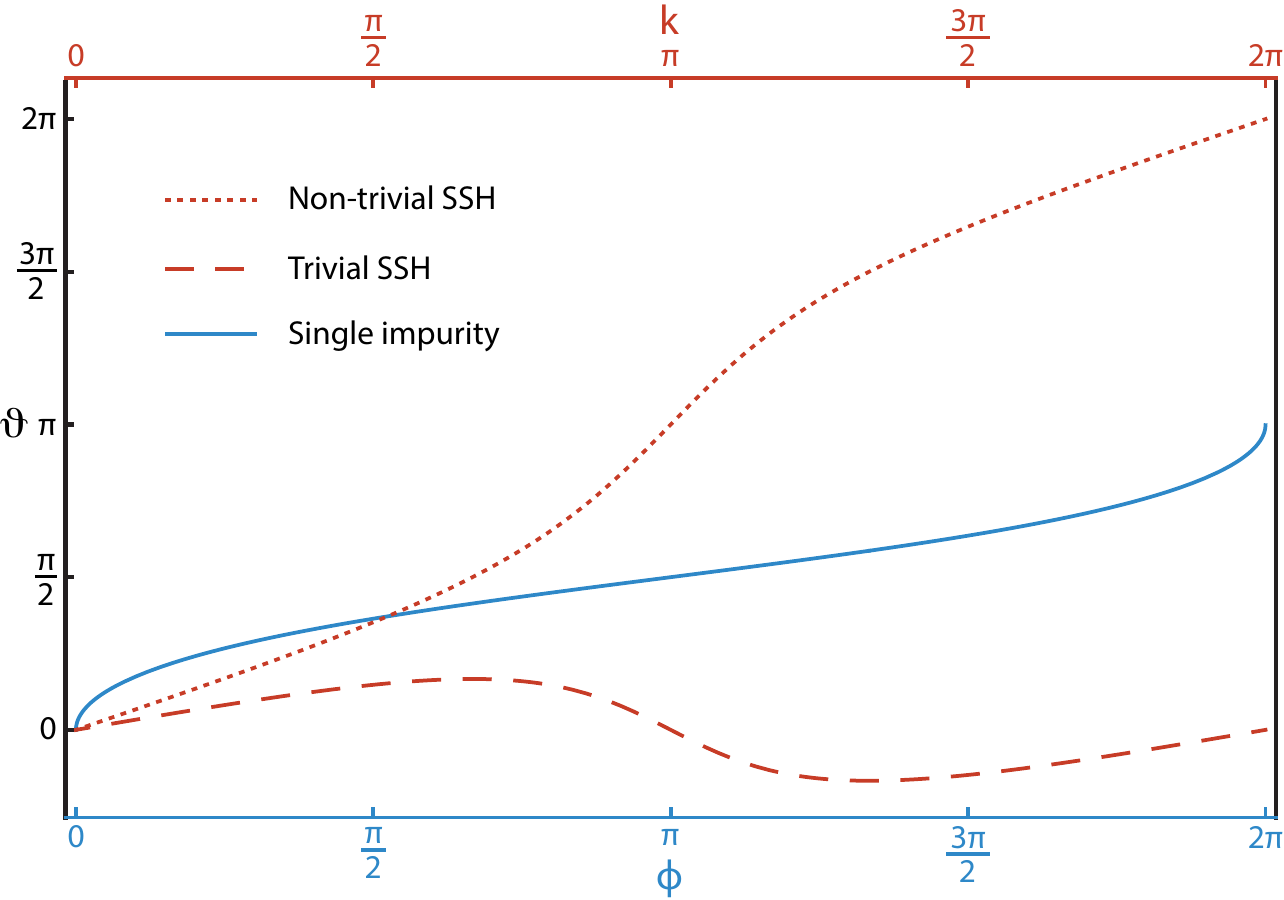} 
		\par\end{centering}
	\caption{\textbf{Phase of impurity state.} Phase $\vartheta$ of the lowest energy eigenstate as a function of the magnetic flux $\phi$ (momentum $k$) for the single-impurity model (SSH model). 
		The color of each curve matches the color of the corresponding horizontal axis. 
		For the SSH model, the ratio $t_1/t_2=2$ ($t_1/t_2=0.5$) was used for the topologically (non-)trivial case.}
	\label{fig:phases}
\end{figure}
From the profile of $\vartheta(k)$ shown in Fig.~\ref{fig:phases}, which has been experimentally confirmed recently in different platforms \cite{Li2023,Neder2024}, one readily obtains $\gamma=0(\pi)$ for the (non-)trivial phase.
According to the bulk-boundary correspondence, an open SSH chain will have zero edge states in the trivial $\gamma=0$ phase, and two edge states in the non-trivial $\gamma=\pi$ phase, exponentially localized at opposite ends of the chain (up to hybridization effects stemming from finite size corrections).

Let us turn our attention to the single-impurity case, with the gapped effective two-band spectrum given by the solid curves in Fig.~\ref{fig:diamond}(c).
It is clear that the relevant parameter of integration is now, not the momentum $k$, but rather the reduced magnetic flux $\phi$, as it changes by one flux quantum, that is, from $\phi=0$ to $\phi=2\pi$.
By inserting $\ket{\tilde{\varphi}_-}=\ket{\tilde{\varphi}_-(\phi)}$, given in (\ref{eq:eigssingleimp}), in (\ref{eq:zak}) leads to (\ref{eq:zakfinal}) with $k\to\phi$.
Therefore, the Zak phase depends on the total variation of $\vartheta(\phi)$ in (\ref{eq:vartheta}) over one flux quantum.
The blue solid curve in Fig.~\ref{fig:phases} shows that $\Delta\vartheta=\pi$, leading to a Zak phase $\pi$-quantized to a half-integer value, $\gamma=\frac{\pi}{2}$.
Alternatively, after rewriting the effective impurity matrix in (\ref{eq:hamiltsingleimp}) as
\begin{equation}
	\mathbf{\tilde{V}}_{_{FB}}=\frac{\epsilon}{4}\sigma_0 +
	\begin{pmatrix}
		0&h(\phi)
		\\
		h^\ast(\phi)&0
	\end{pmatrix},
\label{eq:hamiltsingleimpchiral}
\end{equation} 
where $h(\phi)=\frac{\epsilon}{4}e^{i\vartheta(\phi)}$, one can compute the winding number \cite{Asboth2016,Maffei2018} as
\begin{equation}
	W=\frac{1}{2\pi i}\int_{0}^{2\pi}d\phi\frac{d}{d\phi}\log h(\phi)=\frac{\Delta\vartheta}{2\pi}=\frac{\gamma}{\pi}=\frac{1}{2},
	\label{eq:winding}
\end{equation}
which similarly produces a half-integer value, meaning that $\phi$ needs to change adiabatically by two full cycles to produce one complete winding of $h(\phi)$ around the origin in the complex plane.
Furthermore, from Fig.~\ref{fig:phases} and the form of the eigenvectors in (\ref{eq:eigssingleimp}), it can be seen that they are $4\pi$-periodic in $\phi$ [since both energy bands are flat, see (\ref{eq:enersingle}), the $4\pi$-periodicity is not explicitly manifested there], and also that the eigenvectors exchange in multiples of $2\pi$ as
\begin{equation}
	\ket{\tilde{\varphi}_\pm(\phi)}=\ket{\tilde{\varphi}_\mp(\phi+2\pi)}=\ket{\tilde{\varphi}_\pm(\phi+4\pi)}.
	\label{eq:eigsexchange}
\end{equation}
This behavior is similar to that observed in chiral-symmetric non-Hermitian 1D models where, due to the fact that each exceptional point (EP) \cite{Bergholtz2021} carries a topological charge of $\pm 1/2$, the absolute value of the winding number, $\vert W\vert$, corresponds to half the number of times each EP is encircled as the momentum $k$ traverses the BZ (assuming the wrapping around all EPs follows the same circulation, such that all individual contributions to $W$ have the same sign).
In particular, if a single EP is encircled once then one has $\vert W\vert=1/2$ \cite{Lee2016,Viyuela2016,Leykam2017} and, analogously, the eigenstates of both bands (assuming a two-band model) are also exchanged after traversing the BZ \cite{Lee2016,Shen2018}.
This non-Hermitian $\vert W\vert=1/2$ case has been associated with the presence of a single edge state under open boundary conditions (OBC), and each $1/2$ increment in $\vert W\vert$ has been further associated with the appearance of an extra edge state \cite{Yin2018}.
Concerning the effective model for the single-impurity case studied in this section, it corresponds, to the best of our knowledge, to the first example of a Hermitian model with a half-integer winding number.

Regarding the Zak phase, on the other hand, the other known example of a 1D Hermitian model where it yields a half-integer value at its energy gaps is the diamond chain with $\pi$-flux per plaquette \cite{Kremer2020,Marques2021} [where $W$ is not well-defined since the off-diagonal block of the Hamiltonian in its chiral representation, corresponding here to the scalar $h(\phi)$ in (\ref{eq:hamiltsingleimpchiral}), is not a square matrix].
The $\gamma=\frac{\pi}{2}$ value for the finite-energy bands of this model is a consequence of having a noncentered inversion-axis within the unit cell \cite{Marques2019}, such that the inversion operator picks up a $k$-dependence over one of the sublattices.
Accordingly, a modified bulk-boundary correspondence is manifested by this system, one that matches a half-quantized Zak phase at a given energy gap with the presence of a single edge state at that gap, when open boundaries are considered \cite{Kremer2020,Marques2021,Pelegri2019,Pelegri2019b}.
\begin{figure}[ht]
	\begin{centering}
		\includegraphics[width=0.95 \columnwidth]{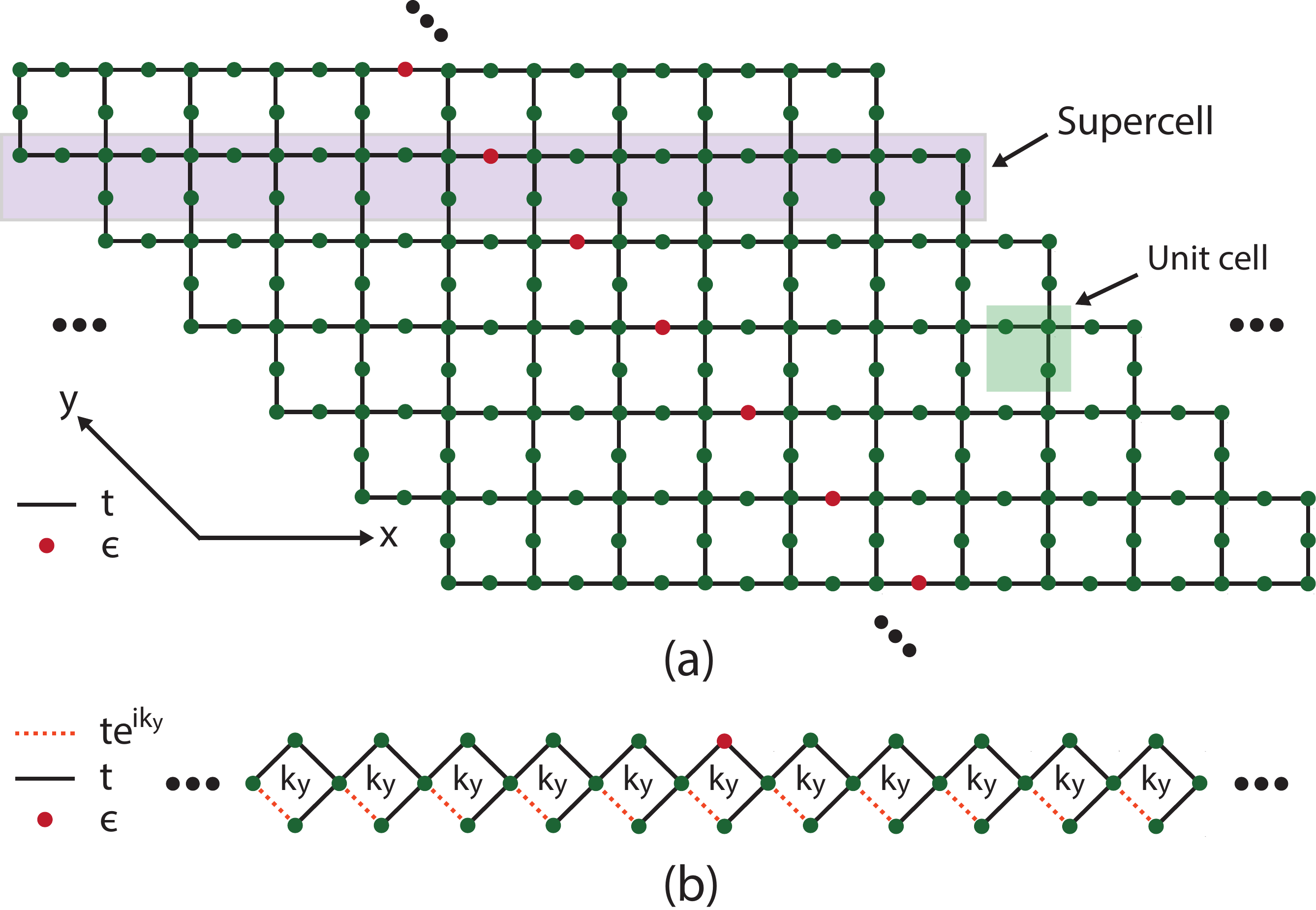} 
		\par\end{centering}
	\caption{\textbf{Mapping between Lieb lattice and diamond chain.} (a) Lieb lattice with a mid-length line of onsite impurity potentials $\epsilon$ at every other site (red dots) along the $y$-direction. All hopping terms are given by $t$.
		(b) Diamond chain obtained from Fourier transforming the lattice in (a) along the periodic $y$-direction, using the indicated supercell. The $y$-momentum $k_y$ becomes the equivalent of a flux per plaquette. The dashed hoppings pick up a positive $k_y$ phase in the clockwise direction within each plaquette.}
	\label{fig:lieb}
\end{figure}

From the examples mentioned above, the same pattern seems to emerge both for Hermitian and non-Hermitian models, namely that a half-integer bulk topological invariant is connected to the presence of a single edge state under OBC. 
The question now becomes: given that the same half-integer invariants (winding number and Zak phase) are observed for our single-impurity case, is there a physical signature of this non-trivial behavior?
Recall that the ``bulk'' bands in Fig.~\ref{fig:diamond}(c) are defined over the reduced flux $\phi$, and not over a momentum variable $k$.
Therefore, the concept of OBC does not directly apply here.
Note that the bulk-boundary correspondence relates quantized bulk invariants defined for the momentum-space with the presence of edge states under OBC in the complementary real-space, since position and momentum are conjugate variables. The conjugate variable of the (reduced) magnetic flux is the (reduced) electric charge, a central relation in the study of quantum electrical circuits \cite{Li1996}. However, and for our purposes, it is not clear what the meaning of OBC in a ``charge-space'' could be, if any, and what would correspond to an edge state in this quantized charge-space.
To circumvent this limitation, we adopt a strategy based on an exact mapping as follows.
First, we consider the modified Lieb lattice depicted in Fig.~\ref{fig:lieb}(a), with a line of impurity sites at every other site along the $y$-direction (red sites), positioned at the center of the $x$ coordinate.
Then, assuming PBC (OBC) in the $y$-direction ($x$-direction), we Fourier transform the lattice along the $y$-direction (with lattice constant set to unity) using the supercell represented in Fig.~\ref{fig:lieb}(a), obtaining the open diamond chain in Fig.~\ref{fig:lieb}(b).
As a result, this chain becomes the same as the single-impurity case of Fig.~\ref{fig:chain}(a), with $k_y$ taking the place of $\phi$, and up to an inconsequential gauge transformation that accumulates all the $k_y$ ``flux'' in a single hopping term within each plaquette.
Therefore, for a small value of $\epsilon$ we expect the effective low-energy subspace to behave as in Fig.~\ref{fig:diamond}(c), with $\phi\to k_y$.
The same half integer invariants $\gamma$ and $W$ in (\ref{eq:zakfinal}) and (\ref{eq:winding}), respectively, are obtained when integrating now over the Brillouin zone in $k_y$.
\begin{figure*}[ht]
	\begin{centering}
		\includegraphics[width=0.9 \textwidth]{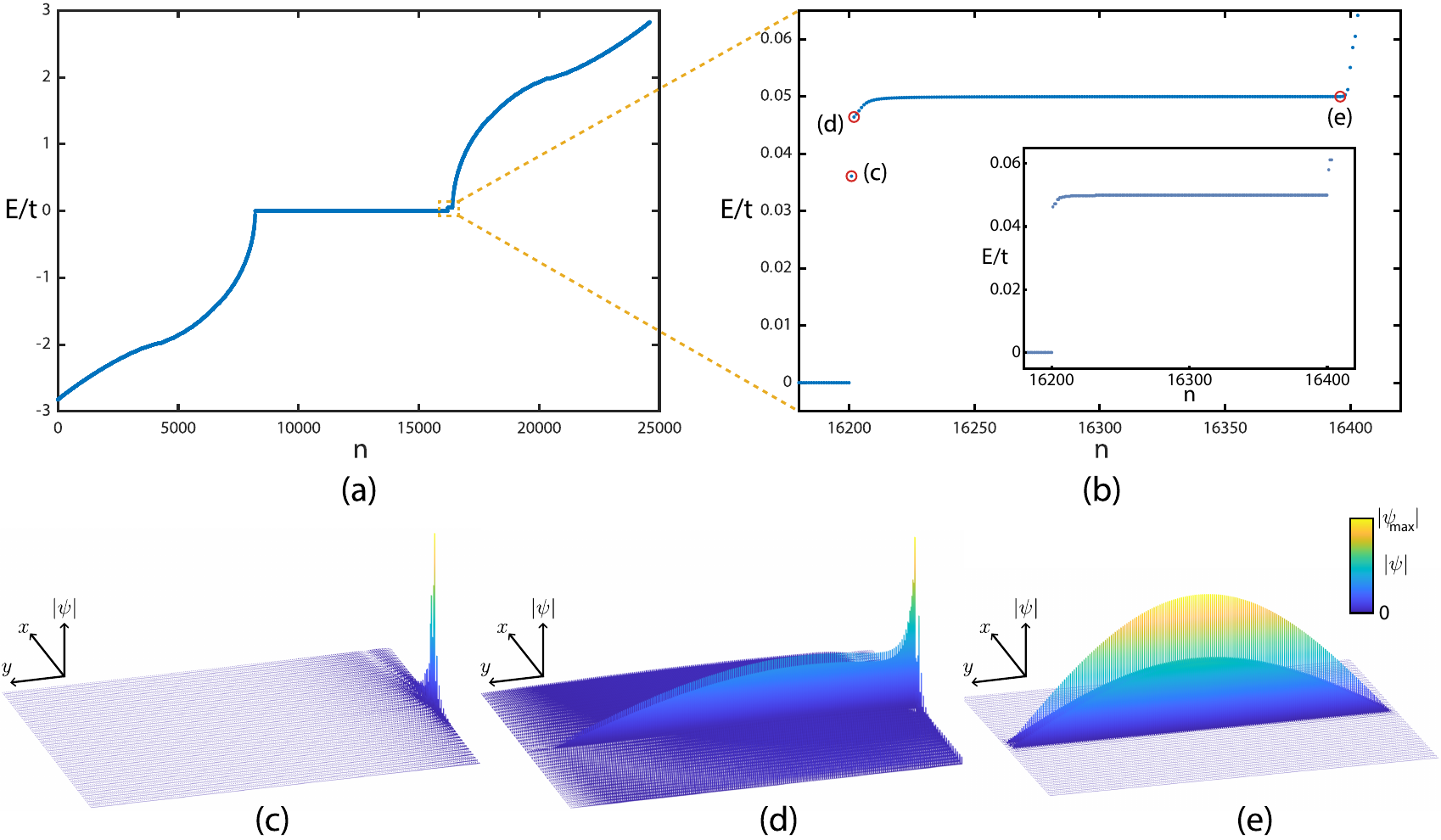} 
		\par\end{centering}
	\caption{\textbf{Lieb lattice with central impurity line.} (a) Energy spectrum, with $t$ as the energy unit, of the impurity decorated open Lieb lattice, with $N_x=41$ ($N_y=200$) unit cells in the $x$ ($y$) direction and impurities $\epsilon=0.1t$ placed along unit cell 21 in $x$ according to the pattern in Fig.~\ref{fig:lieb}(a).
	(b) Zoomed energy spectrum of the region delimited by the orange dashed box in (a).
	The inset shows the energy spectrum of the zoomed region for the same Lieb lattice, only with PBC in the $y$-direction.
(c)-(e) Depiction of the amplitude distribution, normalized by the maximum value for each case, of the eigenstate with index (c) $n=16201$, (d) $n=16202$, and (e) $n=16396$, with energies given by the corresponding encircled states in (b).}
	\label{fig:liebspectrum}
\end{figure*}
In order to test if this translates into the appearance of a gapped edge state in real-space, we impose OBC along the $y$-direction also, \textit{i.e.}, we consider the Lieb lattice in Fig.~\ref{fig:lieb}(a) with OBC for both directions.
In Fig.~\ref{fig:liebspectrum}(a), we plot the energy spectrum for a chain $N_x=41$ ($N_y=200$) unit cells in the $x$ ($y$) direction, with impurities $\epsilon=0.1t$ placed along unit cell 21 in $x$ according to the pattern in Fig.~\ref{fig:lieb}(a).
The overall features of the spectrum in Fig.~\ref{fig:liebspectrum}(a) are the expected ones for a Lieb lattice, namely a gapless system of two continua of dispersive states above and below the localized states of the zero-energy flat band.
However, a closer look at the states with energies around $0\leq E\lesssim\frac{\epsilon}{2}$ [notice how these limits correspond to the energies of the flat bands of the effective system in Fig.~\ref{fig:diamond}(c)], shown in the zoomed region of Fig.~\ref{fig:liebspectrum}(b), reveals the presence of an energy gap.
As expected, an isolated edge state is seen to appear within this gap, with the spatial profile depicted in Fig.~\ref{fig:liebspectrum}(c).
Conversely, when PBC are imposed in the $y$-direction, the in-gap state disappears, as shown in the inset of Fig.~\ref{fig:liebspectrum}(b), which further corroborates its topological origin.

The presence of an energy gap was not guaranteed \textit{a priori} since, loosely speaking, the spectrum sweeps all the $k_y$ values allowed by the OBC, and the gap between the dispersive and flat bands closes as $k_y\to 0$ \cite{Mizoguchi2024}.
Moreover, contrary to the analytical results for the effective model, the numerical results of Fig.~\ref{fig:diamond}(c) also show that the gap between the two bands closes at the edges (see the descending blue dots tending towards the red ones near $\phi\to k_y=0,2\pi$).
But since this gap closing region is very sharp, getting narrower as $\epsilon\to 0$, few solutions are expected to appear in this region under OBC.
Indeed, some states at the left edge of the top band in Fig.~\ref{fig:liebspectrum}(b) are bending towards the zero-energy band, while nonetheless keeping close to the top flat band at $E=\frac{\epsilon}{2}=0.05$, such that the gap remains open and the edge state fully gapped.
In Fig.~\ref{fig:liebspectrum}(d), we show the spatial profile of the leftmost state of this subset.
It can be seen that, while consisting mostly of a bulk state along the impurity line, there is still a finite weight on the region of the right edge state, which gradually shrinks as the state index increases, eventually vanishing once the flat band states at $E=0.05$ are reached [see the profile of the state in Fig.~\ref{fig:liebspectrum}(e)].
We attribute the right edge component for this subset of states to a perturbative hybridization with the gapped edge state that further sharpens their energy bending.

The results of this exact mapping provide another example where the half-integer topological bulk invariants, $\gamma=\pi W=\frac{\pi}{2}$, have a correspondence with the presence of a single gapped edge state under OBC.
In turn, this provides indirect evidence that the topologically non-trivial nature of the single-impurity case studied in this section can be related to specific physical manifestations.
We note that this bulk-boundary correspondence should also be observed upon switching the single impurity from the $B$ site of the midchain plaquette, as in Fig.~\ref{fig:lieb}(b), to its $C$ site.
	In terms of the Lieb lattice in Fig.~\ref{fig:lieb}(a), this corresponds to shifting the impurities along the impurity line, exchanging its red and green sites.
	As we detail in Supplementary Note~1, this shift leads to a physically distinct configuration that produces small changes to the low-energy spectrum in Fig.~\ref{fig:liebspectrum}(b).
	However, as we show in Supp. Fig. 1, the validity of this mapping is not affected, as a single in-gap state is still present.

\section*{Discussion}
\label{sec:discussion} 

We studied a diamond chain with a gapped FB in its energy spectrum upon introducing a finite magnetic flux per plaquette, whose states can be described as CLSs spanning a non-orthogonal basis subspace.
After outlining the mathematical formalism required for the analytical treatment of such non-orthogonal bases, we applied it to the general solutions of the impurity states that are lifted from the FB by small local potentials placed at the midchain plaquette.
For the two equal, two opposite and single-impurity cases, the numerical results were shown to closely follow the analytical predictions, except for flux values very close to the critical gap closing point between flat and dispersive bands, corresponding to the expected breakdown region of the theory.

Disorder studies were conducted for the case of equal impurities.
By applying small disorder following a uniform distribution to the diamond chain, the energy histogram of both impurity states was seen to follow a normal distribution for a wide range of flux values.
These findings were connected to the CLT, such that, from relating the standard deviations of the normal and uniform distributions, an effective number of sites involved in a disorder averaging effect \cite{Munoz2018} was computed for each impurity state.
We discussed how this effect translates into an enhanced robustness of the states to diagonal disorder, which is one of the most common kinds of disorder present in real systems.

A topological characterization of the impurity levels was carried out, where the flux was interpreted as a synthetic momentum.
While the equal and opposite impurity cases were characterized as topologically trivial, the single-impurity case manifested a peculiar type of non-trivial topology, namely one defined by half-integer invariants such as the Zak phase and winding number.
The latter corresponds, as far as we know, to the first of its kind in the context of Hermitian models.
Given that the invariants are computed by considering the reduced magnetic flux as the ``bulk'' variable of integration, and the variation of the flux by a flux quantum as the ``Brillouin zone'', the bulk-boundary correspondence cannot be directly checked.
However, an indirect way of circumventing this limitation was formulated, based on an exact mapping between the diamond chain with a single impurity and a two-dimensional Lieb lattice with a central line of impurities.
The mapped lattice displays the same low-energy effective impurity bulk bands, classified by the same half-integer invariants.
For fully open boundaries, its energy spectrum reveals the presence of a single edge state within the gap of this low-energy subspace, providing an indirect evidence of the unconventional bulk-boundary correspondence associated with the topology of the single-impurity case.

Regarding the realization of the model studied here, shown in Fig.~\ref{fig:chain}(a), we highlight two platforms. 
One is electrical circuits \cite{Lee2018}, whose flexibility and plasticity allows one to recreate most tight-binding models simply by coupling nodes with capacitors or inductors in the desired geometry \cite{Chase2024}.
By incorporating negative impedance converters with current inversion (INICs) \cite{Liu2021a,Zou2021} in phase-control units, it has been shown \cite{Ezawa2020b} that arbitrary phase factors can be associated to the couplings, enabling the generation of a synthetic flux at each plaquette, as required by the target model of Fig.~\ref{fig:chain}(a).
The impurity potentials at the central nodes can be simulated by adding extra capacitors, with small capacitance values, to the grounding schemes of these nodes. 
Preliminary numerical studies should be undertaken to determine the optimal parameters of the circuit since, on the one hand, the admittance elements of the impurity nodes should be small, compared to the admittance gap between flat and dispersive bands, and, on the other hand, they should be sufficiently higher than the tolerance of the other components, so as to not have its effects washed way by this inherent disorder.
As such, very low tolerance components should be considered.
After feeding the circuit with an alternated current with the eigenfrequency of the impurity states, standard voltage and two-point impedance measurements can be used to detect them.

The other strong possibility consists in implementing the diamond chain in photonic lattices.
Three different options can be considered: 
(i) photonic waveguide arrays, where the synthetic flux can be generated by an appropriate bending of the lattice along its longitudinal direction \cite{Mukherjee2018}, while the impurity potentials are controlled by the index of refraction of the corresponding waveguides (which in turn depends on the difference in lasing speed that is used to write them \cite{Szameit2010});
(ii) microring resonator arrays, where the synthetic flux can be controlled by displacing or modifying the optical path of the auxiliary rings used to effectively couple the main ones \cite{Viedma2024,Chen2024,Flower2024}, and the impurities can be introduced by locally perturbing the resonance condition of the main lattice rings \cite{Longhi2015} (through slightly modifying their size, width or refractive index);
(iii) arrays of coupled multiorbital waveguides \cite{Schulz2022,Aravena2022,Jiang2023,Mazanov2023} guiding staggered orbital angular momentum (OAM) modes of order $l=0,\pm 1$ where, in analogy with systems of ultracold atoms \cite{Pelegri2019c,Nicolau2023a,Nicolau2024}, the synthetic flux can be produced by geometrical manipulation of the waveguide pattern of the excited OAM modes \cite{Jorg2020,Wang2024}, while local potentials can again be introduced via changes to their radii or refractive index.

Our work provides a conceptual frame within which future studies on systems with non-orthogonal bases can be carried out.
In particular, we discuss in greater detail in a companion piece to this paper \cite{Viedma2024b} how one can, using the implementation platform (iii) above, and by decorating the diamond chain with impurities placed at regular intervals, extract impurity subspaces realizing different effective models from the FB, including topological ones hosting edge states.

\section*{Data Availability}
\label{sec:data}
Data sharing not applicable to this article as no datasets were generated or analysed during the current study.

\section*{Acknowledgments}
\label{sec:acknowledgments}

A.M.M. and R.G.D. developed their work within the scope of the Portuguese Institute for Nanostructures, Nanomodelling, and Nanofabrication (i3N) Projects No. UIDB/50025/2020, No. UIDP/50025/2020, and No. LA/P/0037/2020, financed by national funds through the Funda\c{c}\~ao para a Ci\^encia e Tecnologia (FCT) and the Minist\'erio da Educa\c{c}\~ao e Ci\^encia
(MEC) of Portugal. 
A.M.M. acknowledges financial support from i3N through the work Contract No.~CDL-CTTRI-46-SGRH/2022, and from FCT through the work Contract No.~CDL-CTTRI-91-SGRH/2024. 
D.V. and V.A. acknowledge financial support from the Spanish State Research Agency AEI (contract No. PID2020-118153GBI00/AEI/10.13039/501100011033) and Generalitat de Catalunya (Contract No. SGR2021-00138).

\section*{Author contributions}
\label{sec:author}

A.M.M. and D.V. developed the concept, analyzed the results and compared them with theoretical predictions. D.V. performed the numerical simulations. R.G.D. and A.M.M. developed the theoretical framework. A.M.M. wrote a first draft of the manuscript. V.A. and R.G.D. supervised the project. All authors discussed the results, read and revised the manuscript.

\section*{Competing Interests}
\label{sec:compint}
The authors declare no competing interests.


\bibliography{impurity}

\appendix

\section*{Supplementary Note 1: Comments on the single-impurity case}
\label{app:singleimp}

The Hamiltonian of the diamond chain in Fig.~\ref{fig:chain}(a) can be parametrized as $H=H(\phi,\epsilon_{_B},\epsilon_{_C})$.
For the general case, a $y$-mirror quasi-symmetry, with the mirror axis going through all the spinal $A$ sites, can be defined as
\begin{equation}
	M_yH(\phi,\epsilon_{_B},\epsilon_{_C})M_y^{-1}=H(-\phi,\epsilon_{_C},\epsilon_{_B}),
	\label{eq:ymirror}
\end{equation}
where $M_y$ is the $y$-mirror operator.
Under this quasi-symmetry, the flux is reversed and the impurities exchange their positions, such that it only becomes an exact symmetry for the equal impurities case $\epsilon_{_B}=\epsilon_{_C}$.
On the other hand, under time-reversal, the system behaves as
\begin{equation}
	KH(\phi,\epsilon_{_B},\epsilon_{_C})K^{-1}=H(-\phi,\epsilon_{_B},\epsilon_{_C}),
	\label{eq:timereversal}
\end{equation}
where $K$ is the conjugation operator obeying $KK^{-1}=1$.
The combination of $y$-mirror and time-reversal operations yields
\begin{equation}
	KM_yH(\phi,\epsilon_{_B},\epsilon_{_C})M_y^{-1}K^{-1}=H(\phi,\epsilon_{_C},\epsilon_{_B}).
	\label{eq:ymirrortime}
\end{equation}
For a general eigenstate $\ket{\psi_n}$ of $H$, the Schr\"odinger equation has the form
\begin{equation}
	H(\phi,\epsilon_{_B},\epsilon_{_C})\ket{\psi_n}=E_n\ket{\psi_n}.
	\label{eq:schrodinger}
\end{equation}
Applying $KM_y$ from the left to both sides of (\ref{eq:schrodinger}) leads to
\begin{equation}
	H(\phi,\epsilon_{_C},\epsilon_{_B})KM_y\ket{\psi_n}=E_nKM_y\ket{\psi_n},
	\label{eq:hamiltkmy}
\end{equation}
where (\ref{eq:ymirrortime}) was used.
By defining $\ket{\psi_n^\prime}:=KM_y\ket{\psi_n}$, (\ref{eq:hamiltkmy}) becomes
\begin{equation}
	H(\phi,\epsilon_{_C},\epsilon_{_B})\ket{\psi_n^\prime}=E_n\ket{\psi_n^\prime},
	\label{eq:schrodinger2}
\end{equation}
which, as could be expected, tells us that the energy spectrum is invariant under an exchange of the impurities.

Through the mapping developed for the single-impurity case in Sec.~\ref{sec:singleimp}C, we arrived at the substitution $\phi\to k_y$. 
The corresponding combined quasi-symmetry acting on the diamond chain depicted in Fig.~\ref{fig:lieb}(b)) becomes, from (\ref{eq:ymirrortime}), and up to a gauge transformation that redistributes the $k_y$ ``flux'' equally between all hopping terms of each plaquette,
\begin{equation}
	KM_yH(k_y,\epsilon,0)M_y^{-1}K^{-1}=H(k_y,0,\epsilon).
\end{equation}
As shown in (\ref{eq:schrodinger2}), $H(k_y,\epsilon,0)$ and $H(k_y,0,\epsilon)$ are isospectral.
However, when OBC are considered in the $y$-direction, and for an integer number of supercells, this isospectrality is lost due to the fact that placing the impurities at the $B$ sites [as in Fig.~\ref{fig:lieb}(a)] or at the $C$ sites [corresponding to exchanging the red and green sites along the impurity line in Fig.~\ref{fig:lieb}(a)] corresponds to two nonequivalent configurations.
This can be understood as follows.
When the clean, impurity-free lattice in Fig.~\ref{fig:lieb}(a) with PBC along the $y$-direction is considered, the $y$-mirror axis is shifted from the center of the supercell \cite{Marques2019,Madail2019}, which induces a $k_y$-dependence in the $y$-mirror operator of the lattice \cite{Yue2024}.
More importantly, if OBC are now imposed along the $y$-direction, the lattice loses the $y$-mirror symmetry, since the top and bottom ends are different: from Fig.~\ref{fig:lieb}(a), it can be checked that, for OBC along the $y$-direction and an integer number of supercells, the top end is a horizontal line of sites, while the bottom end is composed of a horizontal sequence of dangling sites, thus breaking $y$-mirror symmetry (it can only be restored by adding extra sites from an incomplete supercell at one of the edges).
This entails that, when the impurity line is included, the positioning of the impurities at the $B$ or $C$ sites leads to physically distinct configurations.
\begin{figure*}[ht]
	\begin{centering}
		\includegraphics[width=0.9 \textwidth]{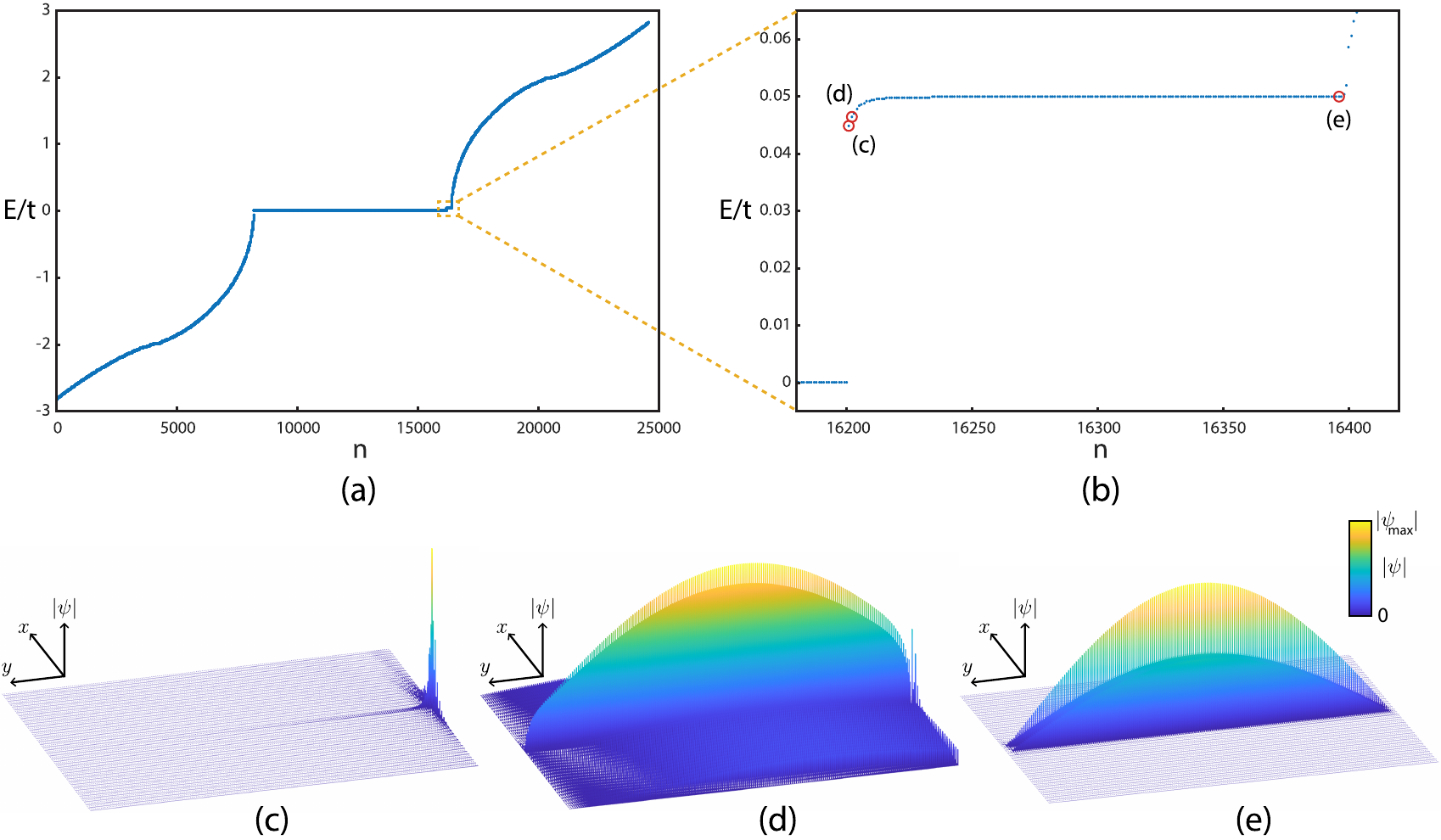} 
		\par\end{centering}
	\caption{\textbf{Lieb lattice with modified central impurity line.} (a) Energy spectrum, with $t$ as the energy unit, of the impurity decorated open Lieb lattice, with $N_x=41$ ($N_y=200$) unit cells in the $x$ ($y$) direction and impurities $\epsilon=0.1t$ placed along unit cell 21 in $x$ as in Fig.~\ref{fig:lieb}(a), but with every impurity shifted by one position along the impurity line (corresponding to exchanging the red and green sites along this line).
		(b) Zoomed energy spectrum of the region delimited by the orange dashed box in (a).
		(c)-(e) Depiction of the amplitude distribution, normalized by the maximum value for each case, of the eigenstate with index (c) $n=16201$, (d) $n=16202$, and (e) $n=16396$, with energies given by the corresponding encircled states in (b).}
	\label{fig:liebspectrum2}
\end{figure*}

The question we want to address here is the following: when changing the impurities from the $B$ sites along the impurity line, as in Fig.~\ref{fig:lieb}(a), to the $C$ sites (by exchanging the red and green sites along this line), do the relevant qualitative features of the low-energy spectrum remain the same?
In other words, does one still observe an energy gap between two flat or nearly-flat bands with a single in-gap edge state, as in Fig.~\ref{fig:liebspectrum}(b), such that the bulk-boundary correspondence is preserved, that is, is the system insensitive to the choice of impurity positioning?
In Supp. Fig.~\ref{fig:liebspectrum2}, we replicate the energy spectrum of the same Lieb lattice with the spectrum of Fig.~\ref{fig:liebspectrum}(a), apart from changing the position of the impurities to the $C$ sites along the impurity line, following the procedure explained above.
As can be seen in the zoomed low-energy spectrum of Supp. Fig.~\ref{fig:liebspectrum2}(b), a single in-gap state is found, albeit much closer in energy to the top bulk band than in Fig.~\ref{fig:liebspectrum}(b).
The spatial profile of this state is depicted in Supp. Fig.~\ref{fig:liebspectrum2}(c), confirming its edge localized nature.
The energy state immediately above this one, depicted in Supp. Fig.~\ref{fig:liebspectrum2}(d), is confirmed to be a bulk state of the impurity line.
Even though this first bulk state of the top band is closer to the in-gap state than in Fig.~\ref{fig:liebspectrum}(b), a weaker hybridization with the edge state than for the state in Fig.~\ref{fig:liebspectrum}(d) is observed.
Finally, the last state of the top band is shown in Supp. Fig.~\ref{fig:liebspectrum2}(e), closely replicating the one in Fig.~\ref{fig:liebspectrum}(e).
In summary, while small quantitative differences are observed between the two choices of impurity positioning, both share the same qualitative features, most notably the presence of a single in-gap state in the relevant low-energy sector of the spectrum.

\end{document}